\documentclass[draft]{agujournal2019}
\usepackage{url} 
\usepackage{soul}
\usepackage{caption}
\usepackage{float}
\usepackage{lineno}
\usepackage[finalnew]{trackchanges}
\soulregister\ref7
\soulregister\cite7
\usepackage{xcolor} 

\draftfalse

\journalname{An edited version of this paper was published by AGU. Copyright (2022) American Geophysical Union.}

\begin{document}

%
%

\title{The Role of Atmospheric Exchange in False-Positive Biosignature Detection}

%
%




\authors{Ryan C. Felton\affil{1,2,3}, Sandra T. Bastelberger\affil{2,4,5}, Kathleen E. Mandt\affil{6}, Adrienn Luspay-Kuti\affil{6}, Thomas J. Fauchez\affil{5,7,8,9}, Shawn D. Domagal-Goldman\affil{5,7}}


\affiliation{1}{Catholic University of America, Washington, DC}
\affiliation{2}{Center for Research and Exploration in Space Science and Technology (CRESST)}
\affiliation{3}{NASA Ames Research Center, Mountain View, CA}
\affiliation{4}{University of Maryland College Park, College Park, MD}
\affiliation{5}{NASA Goddard Sellers Exoplanet Environments Collaboration,
Greenbelt, MD}
\affiliation{6}{John Hopkins Applied Physics Laboratory, Laurel, MD}
\affiliation{7}{NASA Goddard Space Flight Center, Greenbelt, MD}
\affiliation{8}{Goddard Earth Sciences Technology and Research (GESTAR), Universities Space Research Association (USRA), Columbia, MD}
\affiliation{9}{American University, Washington DC, USA}





\correspondingauthor{Ryan Felton}{ryan.c.felton@nasa.gov}




\begin{keypoints}
\item \change{Material exchange}{The transfer of volatiles through atmospheric loss processes} as seen in the Titan-Enceladus system may occur amongst close\add{-}in exoplanets
\item \change{Material exchange amongst terrestrial worlds does not trigger a false-positive biosignature}{We simulate the atmosphere and spectra of TRAPPIST-1 e if it was receiving an external flux of water and oxygen}
\item \change{U}{Our results are important for u}pcoming and future observatories \add{which} must prepare for false-positive biosignatures while searching for life
\end{keypoints} 

%
%

%
%


\begin{abstract}
Saturn's moon Titan receives volatiles into the top of its atmosphere - including atomic oxygen - sourced from cryovolcanoes on Enceladus. Similar types of atmosphere exchange from one body to another, such as O$_2$ and O$_3$ sourced from TRAPPIST-1 d, could be introduced into the upper atmosphere of TRAPPIST-1 e, and might be interpreted as biosignatures. We simulate this potential false positive for life on TRAPPIST-1 e, by applying an external influx of water and oxygen into the top of the atmosphere using a coupled 1-D photochemical-climate model (\textit{Atmos}), to predict atmospheric composition. In addition, synthetic spectral observations are produced with the Planetary Spectrum Generator for the James Webb Space Telescope, Origins Space Telescope, Habitable Exoplanet Observatory and Large UV/Optical/IR Surveyor to test the detectability of abiotic-generated O$_2$ and O$_3$ in the presence of abiotic and biotic surface fluxes of CH$_4$. We determine that the incoming flux of material needed to trigger detection of abiotic O$_2$/O$_3$ by any of these observatories is more than two orders of magnitude (1 $\times$ 10$^{12}$ molecules/cm$^2$/s) above what is physically plausible.
\end{abstract}
\section*{Plain Language Summary}
In the Saturnian system, Enceladus' icy volcanoes spew liquid and gassy material \add{-including water-} into outer space, and some of that material ends up in Titan's atmosphere. \add{This is a prominent phenomenon within our solar system that has been observed in great detail and shows proof of foreign matter exchange being possible between worlds. The simultaneous presence of detectable amounts of methane and oxygen or ozone in an atmosphere is considered strong evidence for the presence of life because in a methane rich atmosphere, oxygen or ozone will be destroyed to the point of being undetectable unless life is present to replenish it and vice-versa.} We explore here whether \change{this type of}{a similar type of} matter exchange, \add{with methane present and} occurring \add{in the} \change{on bodies outside of our solar system}{TRAPPIST-1 system}, could increase\remove{the} oxygen and water \add{abundances to the point of} creating atmospheric signatures that may be mistaken for signs of life on another planet. To probe this question we use computer models of atmospheres and current and next-generation space telescopes. We \remove{ultimately}conclude that when looking for the simultaneous presence of methane and oxygen\change{(ozone)}{/ozone}, this matter exchange will not be mistaken for signs of life.

\section{Introduction} \label{sec:intro}
As we move closer to being able to characterize the atmospheres of terrestrial exoplanets, the ability to identify signs of life and filter out false-positive biosignatures will be of the upmost importance. A biosignature is a detectable property of a planet (or environment) that results from the presence of life on a planet, and whose detection requires the presence of life \cite{lederberg1965signs, lovelock1965physical, marais1999astrobiology, des2008nasa, schwieterman2018exoplanet, meadows2018habitability}. A false-positive biosignature is a means of producing such a detectable property via non-biological processes \cite{harman2018biosignature}. Upcoming observatories have the potential to detect and characterize terrestrial planet atmospheres, and farther-off mission concepts plan to look for biosignatures \change{\cite{gaudi2020habitable,cooray2019origins,luvoir2019luvoir}}{\protect\cite{national2021pathways}}. \citeA{fauchez2020sensitive} have shown that for the TRAPPIST-1 planets in the habitable zone\remove{(HZ)} it will be challenging to detect water or any biosignature gas with the James Webb Space Telescope (JWST). However, \change{CO$_2$}{carbon dioxide} could be detectable and serve as a proxy to infer the presence of an atmosphere. \change{Currently, NASA has four missions being evaluated for the Astro2020 Decadal Survey concepts and three of them (Habitable Exoplanet Observatory, Origins Space Telescope and LUVOIR) will be able to search for biosignatures on potentially habitable exoplanets}{NASA had four mission concepts evaluated for the Astro2020 Decadal Survey and three of them (Habitable Exoplanet Observatory, Origins Space Telescope and Large Ultra-violet Optical Infrared Surveyor - LUVOIR ) will be able to search for biosignatures on potentially habitable exoplanets} \protect\cite{gaudi2020habitable,cooray2019origins,luvoir2019luvoir}. \add{The relevant recommendations of the decadal survery were for a $\sim$6m inscribed diameter Infrared/Optical/Ultraviolet space telescope with a target launch in the first half of the 2040s and a new Probe-class mission type to focus on a far-IR spectroscopy and imaging strategic mission (i.e. Origins) \protect\cite{national2021pathways}.} \add{These recommendations, combined with the successful launch of JWST, motivate the continued study of biosignature and false-positive biosignature simulations in preparation for observational exoplanet atmosphere data in the near-term future.} Prior work has argued that the presence of \change{O$_2$}{oxygen} (and/or \change{O$_3$}{ozone}, which is a photochemical byproduct of O$_2$) in combination with \change{CH$_4$}{methane} in an atmosphere is the most promising biosignature to look for \cite{hitchcock1967life,lovelock1975thermodynamics,domagal2014abiotic,meadows2017reflections}. This classical ``strong" biosignature pair of CH$_4$ and O$_2$\change{(O$_3$)}{/O$_3$} is considered \change{strong}{robust} due to their \change{reaction network}{chemical interaction}. In an atmosphere where O$_2$ is abundant and H$_2$O is present, the hydroxyl radical, OH, is produced which \change{leads to}{promotes} the destruction of CH$_4$ \add{by O$_2$} \cite{levy1971normal}\remove{ and vice-versa.} \add{while a CH$_4$\add{-}rich atmosphere would contain reducing sinks for O$_2$ that rapidly remove it from the atmosphere.} This mutual destruction makes it difficult to maintain detectable levels of O$_3$ in the context of a CH$_4$-rich atmosphere without biological O$_2$ fluxes. \change{t}{Therefore t}he simultaneous presence of CH$_4$ and O$_3$ is \remove{therefore}an indicator of biological production of both CH$_4$ and O$_2$.

\add{Biotic and abiotic sources of atmospheric CH$_4$ and O$_2$ on Earth have been extensively studied and can help constrain plausible flux ranges for this study.} \add{For example CH$_4$ can be produced as a metabolic byproduct of microbial life, and this process is evolutionarily ancient; putative evidence for this metabolism extends all the way back to 3.5 billion years ago \cite{ueno2006evidence}}\remove{CH$_4$ is a known byproduct of life on modern Earth and in Earth's past as is evident by the agricultural industry (cows)}\remove{ and in 3.5 billion year old chert rocks showing evidence of methanogensis during the Archean eon \cite{ueno2006evidence}}.\add{ And while Earth's modern biosphere produces a biological CH$_4$ flux of $\sim$1$\times$10$^{11}$ molecules/cm$^2$/s \protect{\cite{Pavlov2001a}},}\add{ estimates for the Archean range from $\sim$0.3-2.5 times this present-day value \protect{\cite{Kharecha2005}}.} \remove{However,}\change{geological processes, such as serpentinization, can also produce CH$_4$ \cite{berndt1996reduction}}{CH$_4$ is also produced through geological processes, such as serpentinization, but at lower rates \cite{berndt1996reduction}.} \add{This}\remove{which} means there is a range of potential CH$_4$ surface fluxes depending on whether they are produced abiotically or biotically \cite{arney2018organic}\change{. And }{ yet }both\remove{the} \change{biological and abiological fluxes can}{of these forms potentially} lead to accumulation of\remove{potentially} detectable CH$_4$ \cite{krissansen2016detecting}. \add{This would complicate biosignature characterization efforts} \remove{If a CH$_4$ spectral signature was detected,}\add{and require} further context about the planet and host star \remove{is needed, in order}to constrain \change{its}{the magnitude of the} flux and understand its source \cite{krissansen2016detecting,arney2018organic}.

\remove{O$_2$ also has abiotic means of production.} As reviewed in \remove{\cite{meadows2017reflections}}\add{\protect\citeA{meadows2017reflections}}, O$_2$ has characteristics of a potential biosignature, but also has known, albeit specific and detectable, abiotic production mechanisms. These abiotic processes include slower O$_2$ and O$_3$ destruction on water (H$_2$O)-poor planets \cite{gao2015stabilization}, fast O-production via photolysis on CO$_2$-rich \change{ones}{planets} around certain star types \cite{domagal2014abiotic}, and oxidation of the bulk atmosphere, driven by massive hydrogen escape on planets around pre-main sequence M-dwarfs \cite{luger2015extreme}. \change{Previously unmodeled in this prior work on O$_2$ and O$_3$ false positives is the potential for}{Here we consider an additional potential source of O$_2$/O$_3$:} an influx of O from space flowing into the top of \change{an}{the} atmosphere (TOA) to form O$_2$ and O$_3$ abiotically.

Titan provides an example of such an inflow, and its impact on atmospheric chemistry. Titan exhibits rich photochemistry, which produces hydrocarbons, nitriles, and a haze of organic material\add{ \cite{brown2009titan,niemann2010composition}}. Titan's atmosphere is known to contain some oxygen, primarily CO. The origin of oxygen-bearing species in Titan’s atmosphere remained unresolved until the Cassini-Huygens mission. The Cassini Plasma Spectrometer detected a flux of O$^+$ ions into Titan’s atmosphere \cite{hartle2006initial}, suggesting an external source for Titan’s oxygen-bearing species. The active, H$_2$O-rich plumes near Enceladus’ south pole\remove{e.g.} \cite{porco2006cassini} and its H$_2$O torus \cite{hartogh2011direct} provide a likely source. Dissociated and ionized H$_2$O molecules from the plumes are transported through the Saturn system supplying oxygen-bearing ions to Titan’s upper atmosphere. The precipitating O$^+$ ions participate in Titan’s photochemistry, and are likely responsible for the presence of CO, CO$_2$ \cite{horst2008origin,krasnopolsky2009photochemical}, and perhaps even H$_2$O \change{(e.g. \protect\citeA{moreno2012abundance}; \protect\citeA{dobrijevic2014coupling}) in Titan’s atmosphere}{in Titan's atmosphere \protect{\cite{horst2008origin,moreno2012abundance}}}.

\change{Material exchange}{Atmosphere loss and exchange} processes between \change{bodies}{worlds} could be amplified in planetary systems around very active M stars. In such systems, there is the potential for the host star to strip away one planet's atmosphere, leaving it inhospitable. The tight packing of systems like TRAPPIST-1 \cite{gillon2016temperate,gillon2017seven} makes it possible for \change{some}{a larger} portion of that lost atmosphere to be picked up by another close-by planet. \remove{We know that there is material exchange between}Titan and Enceladus\remove{, which} are approximately 10$^6$ km apart, the same order of magnitude as the distance between TRAPPIST-1 d and e. In this work we \change{imagine}{study} \change{the}{a} scenario where an \add{abiotic }exoplanet\remove{with an abiotic atmospheric composition} receives an influx of water or atomic oxygen at the TOA from an outside source. Combined with an abiotic surface flux of CH$_4$, this creates the potential for both CH$_4$ and either O$_2$ or O$_3$ to be detected. If both are simultaneously detectable for reasonable incoming flux rates, this would represent a false-positive for an biosignature that has previously been considered to be ``solid." \remove{Several factors might allow an external flux of O into the atmosphere to create detectable levels of O$_2$ or O$_3$. First, a TOA O (or H$_2$O) flux will be physically separated from the abiotic CH$_4$ flux at the surface due to the large spatial separation (100 km) between the surface and the TOA. This provides the potential that the CH$_4$ and O could mix at rates that are faster than their mutual chemical destruction rates, allowing them to simultaneously accumulate. This means the two chemicals do not interact with each other immediately. And this could be compounded by the lower rates of mixing that occur in the upper layers of an atmosphere, which often have a temperature inversion that hinders vertical mixing \cite{robinson2014common}. If these factors allow O$_2$ or O$_3$ to accumulate, this would occur in the regions of the atmosphere that transit spectroscopy is most sensitive to. However, all of this may not be sufficient to overcome the rapid destruction rates of O$_2$ and O$_3$ in an atmosphere with significant CH$_4$.}To \change{study}{investigate} the potential for incoming \change{material exchange in}{external fluxes into} a terrestrial planet's atmosphere to trigger a CH$_4$-\change{O$_3$(O$_2$)}{O$_2$/O$_3$} biosignature false-positive, we\remove{simulated a planet with an abiotic atmospheric composition} \add{model the atmosphere of an abiotic planet} \change{, with}{exposed to} varying exogenous fluxes of O or H$_2$O \change{. Using a}{using the coupled} photochemical\add{-climate} model, \textit{Atmos} \protect{\cite{arney2016pale}}\remove{ , we predict the atmospheric chemistry profiles for such a world}. We then use the Planetary Spectrum Generator (PSG,\cite{villanueva2018planetary}) to \change{predict}{simulate} transit spectra for those atmospheres, and evaluate the detectability of O$_2$, O$_3$, and CH$_4$.

The paper is broken down into the \change{four following}{following four} sections. Section \ref{sec:method} explains the methods and tools used for the project, Section \ref{sec:result} is a presentation of our main data and most pertinent results, Section \ref{sec:disc} is where we discuss our results and explain what they mean in the context of false-positive biosignatures and finally in Section \ref{sec:conclusion} we tie everything together into our conclusions. 

\section{Methods}\label{sec:method}
\subsection{\add{Coupled} Photochemical\add{-Climate} Modeling - \textit{Atmos}}\label{sec:atmos}
\add{We use the coupled 1D photochemical-climate model, \textit{Atmos}, to simulate the photochemistry and climate of TRAPPIST-1 e.} \remove{\textit{Atmos} is a 1D}\add{The} photochemical model \add{was most recently discussed in}\remove{based on \protect\citeA{kasting1979oxygen}}\remove{ that has been updated over the years to simulate a variety of planets. This project uses the branch of this code most recently discussed in} \add{\protect\citeA{arney2017pale} and \protect\citeA{badhan2019stellar}}\add{. Here we use a version with the updated wavelength grid from \protect\citeA{lincowski2018evolved}}. The model solves the continuity and flux differential equations for the atmospheric species inputted and the solutions correspond to a steady state atmosphere. Boundary conditions of the atmosphere and parameters for the planet and host star are all adjustable by the user. \change{The tunable boundary conditions relevant to this work were: deposition velocity, upward surface flux, downward flux from space into the atmosphere, and fixed mixing ratios.}{For a given atmospheric species, the lower boundary condition can be set to either a fixed deposition velocity, a constant surface flux, a constant flux distributed throughout the troposphere or a constant surface mixing ratio. For the upper boundary, a species can be assigned a constant TOA flux.} \change{The boundary conditions and bulk planet/astrophysical parameters for this project}{In this project, we use}\add{ the planetary parameters for TRAPPIST-1 e from \protect{\citeA{agol2021refining}}} \change{assumed a terrestrial planet with}{and assume an atmosphere with Earth pressure and} abiotic \change{chemical}{surface} boundary conditions driven by redox balance \cite{harman2018abiotic, domagal2014abiotic}\remove{, and the stellar flux of TRAPPIST-1 e, which orbits the M-star TRAPPIST-1, combined with the mass and radius of Earth}. The star and planetary parameters are listed in Table \ref{starplanetfeatures} and a list of \add{the} species with non-zero boundary conditions can be found in Table \ref{speciestable}.\add{ A complete list of all species and boundary conditions is available in the supplemental material found in the Zenodo data repository link in the acknowledgements section at the end of the paper.} These boundary conditions are paired with a photochemical reaction network, originally designed for Archean Earth \cite{arney2016pale} \change{that we have expanded upon with 429 reactions}{that has been expanded to include more reactions and revised with updated reaction rate data}\remove{supplemental table xx}. The \change{host star and}{applied} stellar flux \change{are}{is} based on \change{synthetic}{the semi-empirical PHOENIX model} spectra of TRAPPIST-1 generated by \citeA{peacock2019predicting}\remove{to best represent TRAPPIST-1's stellar energy distribution and TRAPPIST-1 e's total incoming stellar radiation}. We did not generate any appreciable haze since our largest CH$_4$/CO$_2$\remove{mixing} ratio was \change{only 5 $\times$ 10$^{-4}$}{4 $\times$ 10$^{-3}$}, which is \add{two orders of magnitude smaller than}\remove{well below} the 0.1 ratio at which haze is expected to form for Archean Earth conditions \cite{trainer2006organic, arney2017pale}. \add{Although TRAPPIST-1 e is likely to be tidally locked \cite{kasting1993habitable, barnes2017tidal}, we use a 1-D model which does not account for temperature and composition differences between the day and night side.} \remove{We also did not assume the planet was tidally locked}\add{To assess how inhomogeneous build-up of O$_2$ due to  tidal locking could affect the accuracy of our predictions, we perform O$_2$ sensitivity scaling tests with our spectral simulator as described in Section 2.2 and Section 4.2}.\remove{ A complete list of all species and boundary conditions are available in the supplemental material.}

\begin{table}
\caption{Host Star \& \remove{Hybrid Archean Earth/}TRAPPIST-1 e Planet Parameters}
\centering
\begin{tabular}{l c}
\hline
\textbf{Parameter} & \textbf{Value}\\
\hline
  Host star & TRAPPIST-1 \\
  Stellar type & M8V\\
\hline
  Planet radius [R$_{\oplus}$] & \change{1}{0.92}\\
  Surface gravity [m/s$^2$] & \change{9.8}{8.01} \\
  Surface pressure [bar] & 1.013 \\
  Surface temperature [K] & \change{281}{267} \\
  Instellation [F$_{\oplus}$] & \change{0.61}{0.646} \\
\hline
\multicolumn{2}{p{0.9\columnwidth}}{\textbf{Note}. \add{Planetary parameters were taken from TRAPPIST-1 e \protect{\cite{agol2021refining}}.} \add{We assumed Earth-like surface pressure and the surface temperature is an average obtained from coupled photochemistry-climate simulations.} \remove{The planetary parameters are based on Archean Earth while the stellar instellation is TRAPPIST-1 e specific.}} 
\end{tabular}
\label{starplanetfeatures}
\end{table}

\add{To simulate an abiotic atmosphere we} \change{We combined}{use} abiotic boundary conditions from \citeA{harman2018abiotic}\change{, which}{. This approach} assumes that in the absence of a biosphere, volcanic outgassing and reactions at the seafloor drive most of the chemical fluxes to/from the bottom of the atmosphere\remove{, since there is no life present to contribute to fluxes into the atmosphere}. To this set of boundary conditions, we add\remove{ed} a surface CH$_4$ \add{flux} as well as TOA fluxes of either O or H$_2$O. \change{The}{We vary the} CH$_4$, O, and H$_2$O flux values\remove{were varied} \add{(Table \ref{climadatacombos})}, to see how the detectability of CH$_4$, O$_2$, and O$_3$ change\remove{d} in response to these\remove{abiotic} fluxes. \remove{For each case, we set the range of fluxes from ``plausible and ideally demonstrated" on the low end to ``unlikely/implausible abiotically" on the high end. The methane values ranged from ``possible abiotically" to ``likely biotic."} We follow\remove{ed} \citeA{arney2018organic} and institute\remove{d} \add{CH$_4$ }fluxes (molecules/cm$^2$/s) that \change{were}{are} abiotic (1 $\times$ 10$^9$), \change{in-between abiotic and biotic}{ambiguously abiotic or biotic} (1 $\times$ 10$^{10}$) and biotic (6 $\times$ 10$^{10}$). These bins are consistent with probabilistic assessments of abiotic CH$_4$ fluxes from \citeA{krissansen2016detecting}.

The incoming O fluxes range from known examples in the Solar System on the low end to values that are \change{unlikely}{on the threshold of what is likely} to exist in nature on the high end. The lower limit of 1 $\times$ 10$^{6}$ molecules/cm$^2$/s is based on estimated oxygen flux into Titan's atmosphere from \citeA{sittler2009heavy}\remove{. The physically plausible upper limit on our oxygen flux was 1 $\times$ 10$^{10}$ molecules/cm$^2$/s, based on the exoplanet habitability study from \protect{\citeA{garcia2017magnetic}}}\remove{The authors modeled an Earth-twin of Proxima Centauri b, which orbits a red dwarf, to determine if a planet's magnetic field would protect the planet from stellar activity by simulating ionospheric outflow, specifically O$^+$. The calculated O$^+$ flux had a maximum value of $\sim$10$^{10}$ molecules/cm$^2$/s.} \add{and the upper limit is taken from an exoplanet \change{habitabilitiy}{habitability} study by \protect{\citeA{garcia2017magnetic}}. \protect{\citeA{garcia2017magnetic}}}\add{ simulated stellar induced ionospheric outflow of O$^+$ from an Earth-twin of Proxima Centauri b, calculating the maximum O$^+$ escape flux to be $\sim$10$^{10}$ molecules/cm$^2$/s.}
The high end of the range used in our study assume\change{d}{s} maximum rates of atmospheric O loss from one terrestrial planet with perfect transfer of all material to a second planet. This type of ``perfect transfer" is highly unlikely to occur in nature; so these high flux values represent an extreme end-member, to ensure that we bound all plausible incoming O flux values. \add{For the scenario of an influx of water we follow the same pattern of establishing a lower and upper boundary.}

\add{Despite the implausibility of a water molecule surviving the trip from outer space and into an atmosphere wholly intact, we modelled the experiment this way for multiple reasons. First, there is the stoichiometric argument to be made that even if all of the water molecules were dissociated on their way to the top of a planet's atmosphere, the hydrogen atoms should still be present. Second, using water molecules as a representation for oxygen molecule flux into an atmosphere is common practice by Titan photochemical modelers \protect{\cite{hebrard2012neutral}}}.\add{ Finally, by modelling a water flux we are able to account for other external oxygen sources to an atmosphere such as water ice (i.e. cryovolcanism on Enceladus). We set our incoming water flux lower bound at 5 $\times$ 10$^{6}$ molecules/cm$^2$/s based off of the work by \protect\citeA{hebrard2012neutral}.}\remove{The incoming water flux lower bound of 5 $\times$ 10$^{6}$ molecules/cm$^2$/s is used by \protect{\citeA{hebrard2012neutral}}}\remove{ when simulating the incoming O flux to Titan's atmosphere in the form of H$_2$O molecules (instead of O atoms). }\remove{Photolysis of the incoming water can provide the O necessary to build up O$_2$ and O$_3$.} The upper bound water flux is based on estimated water loss on Earth, assuming perfectly efficient transfer of this water to a second planet. Similar to the upper end of our range of O fluxes, this is intentionally beyond the range of what we consider to be likely to occur in nature, as a means of ensuring that we have included all possible incoming H$_2$O flux rates. \change{In both the O and water flux }{For both the O and H$_2$O flux} \add{there is the potential that neither of these scenarios will trigger a simulated O$_2$/O$_3$ signal. If this occurs we want to be able to quantify the fluxes needed to generate a false-positive. To prepare for this possibility we extend our simulations up to 1 $\times$ 10$^{12}$ molecules/cm$^2$/s.} \remove{scenarios we explored flux values as high as 1 $\times$ 10$^{12}$ molecules/cm$^2$/s but only in an attempt to trigger an O$_2$ (O$_3$) signal if the physically plausible upper limits did not themselves allow a signal to form.}

\begin{figure}[H]
\centering
    \includegraphics[trim = 130 85 100 99, clip,scale=0.75]{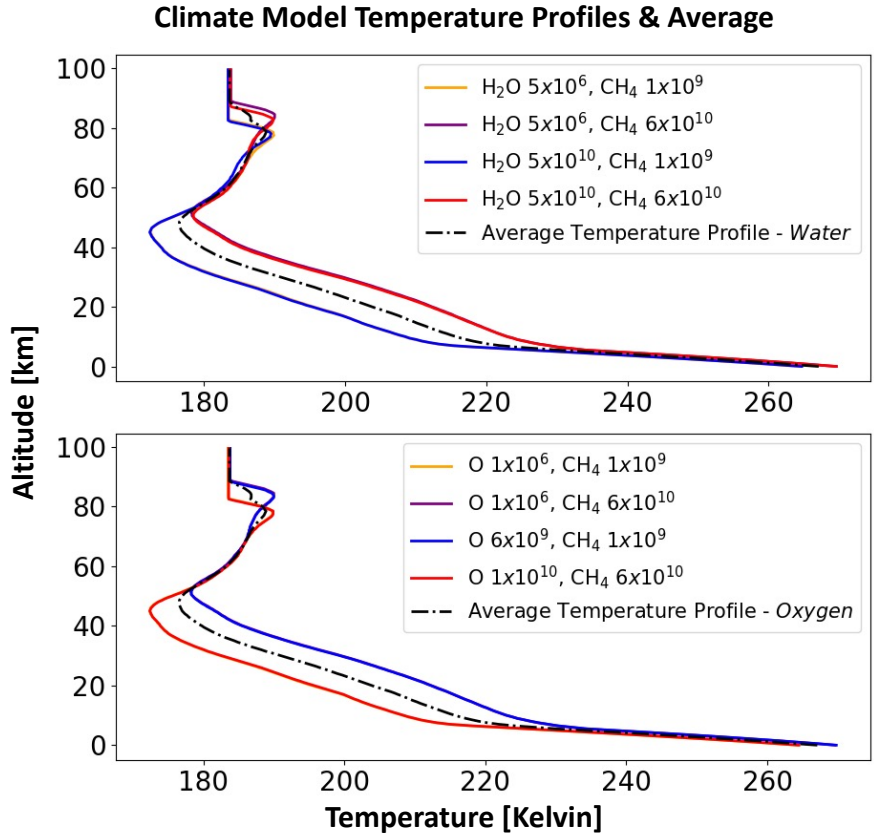}
    \caption{\add{Temperature profiles obtained with the coupled photochemistry-climate model (solid lines) and averaged (dash-dotted lines) temperature profiles for \textit{water} (upper panel) and \textit{oxygen} influx (lower panel). The two averaged temperature profiles are used for all of the photochemical and spectral simulations.}}
    \label{fig:clima_temp_profiles}
\end{figure}

\add{The temperature-pressure profiles we use in this study are obtained by coupling the photochemistry model to the Atmos climate model. Derived from the 1D model created by \protect{\citeA{kasting1986climatic}}}\add{, the version of the climate code used here has undergone many updates \protect{\cite{Kopparapu2013,arney2016pale}}}. \add{The code uses the $\delta$ two-stream approximation \protect{\cite{Toon1989}}} \add{and absorption by the spectrally active gases - H$_2$O, O$_2$, O$_3$, CH$_4$ and C$_2$H$_6$ - is calculated based on the correlated-$k$ method. The model's H$_2$O and CO$_2$ $k$-coefficient have recently been recomputed using the HITRAN2016 database \protect{\cite{GORDON2015}}.}\add{ We use the Mega-MUSCLES TRAPPIST-1 spectral energy distribution from \protect{\citeA{wilson2021mega}}} \add{for the climate model. In coupled mode, the photochemistry model passes the vertical mixing ratio profiles of the spectrally active gases and planetary parameters to the climate model. These are used by the climate model to compute the temperature-pressure profile and tropospheric water abundance which are in turn passed to the photochemistry model. The models iterate back and forth until both individual models are converged and a self-consistent atmospheric solution in radiative convective equilibrium is determined. For this study, we assume the tropospheric relative humidity parameterization by \protect{\citeA{Manabe1967}}.} \add{We run eight coupled photochemistry-climate simulations applying O$_2$, H$_2$O and CH$_4$ fluxes within the physically plausible range (Table \ref{climadatacombos}). Subsequently, averaged temperature pressure profiles are computed and used as input for every photochemical simulation, see Fig. \ref{fig:clima_temp_profiles}. Note that we are unable to arrive at a coupled solution for the boundary condition O flux 1 $\times$ 10$^{10}$ molecules/cm$^2$/s, CH$_4$ flux 1 $\times$ 10$^9$ molecules/cm$^2$/s and instead use the O flux 6 $\times$ 10$^9$ molecules/cm$^2$/s.}

\begin{table}
\caption{Photochemical Model Boundary Conditions}
\centering
\begin{tabular}{l c r}
\hline
Species & Boundary Type & Value \\
\hline
 \textbf{O} & $\nu_{dep}$, \textbf{downward flux} & 1, \textbf{1 $\times$ 10$^6$ - 1 $\times$ 10$^{10}$}\\
 O$_2$ & $\nu_{dep}$ & 1 $\times$ 10$^{-4}$\\
 \textbf{H$_2$O} & $\nu_{dep}$, \textbf{downward flux} & 0, \textbf{5 $\times$ 10$^6$ - 5 $\times$ 10$^{10}$}  \\
 H & $\nu_{dep}$ & 1\\
 OH & $\nu_{dep}$ & 1\\
 HO$_2$ & $\nu_{dep}$ & 1\\
 H$_2$O$_2$ & $\nu_{dep}$ & 0.5\\
 H$_2$ & $\nu_{dep}$, upward flux & \change{6.81 $\times$ 10$^{-4}$}{6.81 $\times$ 10$^{-6}$}, 1 $\times$ 10$^{10}$\\
 CO & $\nu_{dep}$ & 1 $\times$ 10$^{-8}$\\
 HCO & $\nu_{dep}$ & 1\\
 H$_2$CO & $\nu_{dep}$ & 0.1\\
\textbf{CH$_4$} & \textbf{upward flux} & \textbf{1 $\times$ 10$^9$ - 6 $\times$ 10$^{10}$}\\
 CH$_3$ & $\nu_{dep}$ & 1\\
 C$_2$H$_6$ & $\nu_{dep}$ & 1 $\times$ 10$^{-5}$\\
 NO & $\nu_{dep}$ & 3 $\times$ 10$^{-4}$\\
 NO$_2$ & $\nu_{dep}$ & 3 $\times$ 10$^{-3}$\\
 HNO & $\nu_{dep}$ & 1\\
 N & downward flux & 1 $\times$ 10$^8$ \\
 H$_2$S & $\nu_{dep}$, upward flux & 0.015, 3 $\times$ 10$^8$ \\
 HSO & $\nu_{dep}$ & 1\\
 H$_2$SO$_4$ & $\nu_{dep}$ & \change{1}{0.2}\\
 SO & $\nu_{dep}$ & 3 $\times$ 10$^{-4}$\\
 CO$_2$ & fixed mixing ratio & 0.02\\
 \add{S} & \add{$\nu_{dep}$} & \add{1}\\
 \add{HS} & \add{$\nu_{dep}$} & \add{\protect{3 $\times$ 10$^{-3}$}}\\
 \add{SO$_2$} & \add{$\nu_{dep}$, upward flux} & \add{1}, \add{\protect{3 $\times$ 10$^9$}} \\
 \add{SO$_4$ Aerosol} & \add{$\nu_{dep}$} & \add{\protect{1 $\times$ 10$^{-2}$}}\\
 \add{S$_8$ Aerosol} & \add{$\nu_{dep}$} & \add{\protect{1 $\times$ 10$^{-2}$}}\\
 \add{C$_4$H$_2$ Aerosol} & \add{$\nu_{dep}$} & \add{\protect{1 $\times$ 10$^{-2}$}}\\
 \add{C$_5$H$_4$ Aerosol} & \add{$\nu_{dep}$} & \add{\protect{1 $\times$ 10$^{-2}$}}\\\\
\hline
 HNO$_2$ & short-lived\\
 O($^1$D) & short-lived\\
 CH$_2^1$ & short-lived\\
 C & short-lived\\
 SO$_2^1$ & short-lived\\
 SO$_2^3$ & short-lived\\
 HSO$_3$ & short-lived\\
 OCS$_2$ & short-lived\\
 CS$_2^\ast$ & short-lived\\
 \hline
\multicolumn{3}{p{0.80\columnwidth}}{\textbf{Note}. \remove{We only list here the species} \add{Species} with non-zero boundary conditions used in the photochemical mode\add{l}. $\nu_{dep}$ is the deposition velocity (cm/s) for a species, downward flux (molecules/cm$^2$/s) is relative to the TOA, i.e. material entering the TOA from outer space, and upward flux is relative to the surface, i.e. material outgassing from the surface and moving up into the atmosphere. The multiple flux values for O, H$_2$O, and CH$_4$ correspond to the lower and upper physical limit values we impose. \add{For a full list of boundary conditions, see the input files for our simulations at our Zenodo repository in the Acknowledgements section.}}
\end{tabular}
\label{speciestable}
\end{table}

\begin{table}
\caption{Coupled Incoming Flux Combinations}
\centering
\begin{tabular}{l c c}
\hline
& TOA flux & CH$_4$ flux \\
\hline
1) Oxygen flux & 1 $\times$ 10$^{6}$ & 1 $\times$ 10$^{9}$\\
2) Oxygen flux & 1 $\times$ 10$^{6}$ & 6 $\times$ 10$^{10}$\\
3) Oxygen flux & 1 $\times$ 10$^{10}$ & 6 $\times$ 10$^{10}$\\
4) Oxygen flux & 6 $\times$ 10$^{9}$ & 1 $\times$ 10$^{9}$\\
\hline
\hline
5) Water flux & 5 $\times$ 10$^{6}$ & 1 $\times$ 10$^{9}$\\
6) Water flux & 5 $\times$ 10$^{6}$ & 6 $\times$ 10$^{10}$\\
7) Water flux & 5 $\times$ 10$^{10}$ & 1 $\times$ 10$^{9}$\\
8) Water flux & 5 $\times$ 10$^{10}$ & 6 $\times$ 10$^{10}$\\
\hline
\multicolumn{3}{p{0.61\columnwidth}}{\textbf{Note}. \add{Flux input combinations used in the coupled photochemical-climate model runs.}}
\end{tabular}
\label{climadatacombos}
\end{table}

\remove{\textit{Atmos} generated vertical mixing ratio profiles for all species in the model. The profiles of gases known to have an impact on terrestrial planet spectra - N$_2$, H$_2$O, CH$_4$, C$_2$H$_6$, CO$_2$, O$_2$, O$_3$, CO, and H$_2$CO - were sent as inputs to PSG, which was then used to produce synthetic transit spectra. These gases were chosen due to their high mixing ratios and abundances in the atmosphere and due to their known impact on atmosphere characterization efforts.}

\subsection{Synthetic Spectrum Generator - PSG}\label{sec:psg}
\add{After all photochemical simulations are complete the vertical mixing ratio profiles of gases known to have an impact on terrestrial planet spectra - N$_2$, H$_2$O, CH$_4$, C$_2$H$_6$, CO$_2$, O$_2$, O$_3$, CO, and H$_2$CO - are sent as inputs to PSG, to produce synthetic transit spectra.} PSG is an online radiative transfer tool\remove{ created by} \cite{villanueva2018planetary}\change{. PSG}{ that} uses a combination of radiative transfer models and spectroscopic databases to produce synthetic atmospheric spectra with either line-by-line calculations, the correlated-k method\add{,} or a combination of the two. For this project all spectroscopic data use\change{d}{s} a combination of these two methods. Numerous types of planetary bodies (i.e. planets, exoplanets and comets) can be studied with PSG and we use it to simulate the transmission spectrum of our \change{abiotic TRAPPIST-1 e / Archean-Earth hybrid}{simulated planets} as observed by four unique space-based observatories\remove{. These telescopes are}: JWST, Origins, HabEx and LUVOIR. 

\remove{Since neither \textit{Atmos} nor PSG predict clouds self consistently, given atmosphere properties, }\change{c}{C}louds \change{were}{are} prescribed in\remove{to} PSG based on \remove{the}TRAPPIST-1 e Archean Earth-like global climate model (GCM) simulations \cite{fauchez2019impact} \change{performed with the}{using} LMD-G GCM \citeA{wordsworth2011gliese}. Water and ice clouds \change{were}{are} distributed between the surface and 19 km in the atmosphere profile. Water clouds, with a mass mixing ratio [kg/kg] of water droplets of 2 $\times$ 10$^{-6}$, range\remove{d} from the surface to 12 km and then ice clouds (mass mixing ratio [kg/kg] of ice droplets: 2 $\times$ 10$^{-7}$) \change{were}{are} between 13 and 19 km.\remove{The goal was not to explicitly mimic TRAPPIST-1 e modeled and theorized clouds, but instead to introduce clouds that would have a basic impact on the overall detectability of any synthetic spectral features.} For every simulation with clouds, a clear sky case \change{was}{is} also simulated. 

\remove{We simulated all combinations of cases of CH$_4$, O and H$_2$O fluxes as listed in Section \ref{sec:atmos} and Table \ref{speciestable}.}\add{From the photochemical-climate simulations we use the flux combinations $\#$1, 3, 4, 5, 7 and 8 in Table \ref{climadatacombos}. \#1 and $\#$5 are used to look for simulated detections of CH$_4$ at the lowest abundances, $\#$3 and $\#$8 probe for CH$_4$ and O$_2$/O$_3$ at the highest plausible ranges and $\#$4 and $\#$7 are used for O$_2$ sensitivity scaling tests. Since our simulations do not account for tidal locking and 3D GCM considerations (i.e. day-night side modeling), sensitivity tests on O$_2$ accumulation are performed with PSG. This consists of scaling the O$_2$ mixing ratio by a factor of 0.5, 2 and 10 and then analyzing the resulting spectra for O$_2$ features across every observatory. The scaling is performed uniformly throughout the atmosphere on O$_2$'s mixing ratio; N$_2$'s mixing ratio is set to maintain a total atmospheric pressure of 1 bar.} To determine whether or not a signature \change{was}{is} present and detectable by any of the four observatories a signal-to-noise ratio (S/N) \change{had}{has} to be determined. This \change{was}{is} dependent on tuning the resolving power of the instrument, for a specific signal (i.e. CH$_4$ at 7.72 $\mu$m), so that the resulting signal amplitude, continuum height, synthetic noise \add{error} bars and transmittance spectra provide\remove{d} the highest S/N.\remove{Note that TRAPPIST-1 e will only transit 85 times during JWST's 5.5 year nominal life time.} \change{Note that during JWST's 5.5 year nominal life time TRAPPIST-1 e will only be in the observing zone approximately a third of the time and so the system will be observable for 85 transits.}{Note that due to the successful launch of JWST, the amount of fuel present has been estimated to allow a 20 year lifetime. This is in contrast to the original 5.5 nominal lifetime pre-launch. We round this down to 5 years, and use the Contamination \& Visibility Calculator tool (https://exoctk.stsci.edu/contam\_visibility) to determine TRAPPIST-1 e will only be in the observing zone approximately a third of the time and so the system will be observable for 85 transits. We begin with 85 transits to compute the S/N for all telescope simulations and then an additional 1/$\sqrt{hv}$ scaling, where $hv$ is the number of photons, is applied to calculate 10 and 20 year lifetimes (170 and 340 transits respectively).} \remove{Therefore we use that value of 85 transits as our upper limit to compute the S/N. }An atmosphere profile and observatory are selected and given a resolving power (R) to start with, typically R=100. \change{The generator}{PSG} \change{was}{is} run and the resulting spectra \change{was}{are} checked for CH$_4$, O$_2$ or O$_3$ signals. Non-overlapping features \change{were}{are} identified and the best one \change{was}{is} chosen based on the S/N. The S/N of the feature \change{was}{is} calculated by subtracting the continuum from the features amplitude and then dividing that by the features synthetic noise. This provide\remove{d}\add{s} an S/N value and then the resolving power \change{was}{is} lowered from 100 in increments of 10 until the largest S/N \change{was}{is} found. For all observatories\add{, except JWST,} we \change{did}{do} not include their estimated noise floors and instead follow\remove{ed} the procedure of \citeA{fauchez2019impact}\add{ and} \citeA{Pidhorodetska2020} by assuming a photon limited scenario where the noise is perfect and decreases with 1/$\sqrt{n
}$, with $n$ being the number of transits. \add{When JWST simulations are performed a noise floor of 10ppm is assumed.} The observatory specific instruments and parameters are listed below\remove{:}\add{.} \add{Note that although Origins, HabEx and LUVOIR have direct imaging capabilities we \textit{only} simulate the instruments specific to observing exoplanets in transit.}

\begin{enumerate}
\item \change{The}{Simulated} JWST observations use two instruments: the Mid-Infrared instrument (MIRI) with the Low Resolution Spectroscopy (LRS) and the \change{NIRSpec (Near-Infrared Spectrograph)}{Near-Infrared Spectrograph (NIRSpec)} with the Prism mode. MIRI-LRS observes in the 5-12 $\mu$m range and NIRSpec Prism is in 0.7-5 $\mu$m range. The MIRI-LRS configuration use\remove{d}\add{s} a resolving power of R=50 while NIRSpec Prism \change{was}{is} set to 30 resolving power. MIRI-LRS ha\remove{d}\add{s} its S/N optimized by adjusting the resolving power, just as all of the other instruments \change{did}{do}\add{; }\remove{, but }R=50 \remove{instead of R=30}\change{gave}{gives} the best results. We \change{found}{find} this combination of instrument and transits to give the best resolution of our TRAPPIST-1 e hybrid while other scientists studying TRAPPIST-1 e with PSG \cite{fauchez2019impact, Pidhorodetska2020} have also used MIRI-LRS and NIRSpec Prism for simulating TRAPPIST-1 e transit observations.

\item For the Origins simulations we use\remove{d} the Transit Spectrometer of the Mid-Infrared Imager/Spectrograph/Coronagraph (MISC) Instrument in the 2.8-11 $\mu$m range with a resolving power of 30 \cite{cooray2019origins}.

\item The HabEx simulations use\remove{d} the HabEx Work Horse Camera (HWC) with 30 resolving power in the 0.37-1.8 $\mu$m range \cite{gaudi2020habitable}.

\item The LUVOIR simulations \change{were}{are} performed with the High Definition Imager (HDI) instrument, the block A mirror size (15 m) and a resolving power of 30 in the 0.2-2.5 $\mu$m range \cite{luvoir2019luvoir}. \add{The Astro2020 Decadal Survey recommended a space observatory with a  mirror that has a $\sim$6m inscribed diameter, that would be a combination of the ideas and concepts laid out for LUVOIR and HabEx \protect\cite{national2021pathways}.} \add{However, since our larger LUVOIR-A simulations do not produce a simultaneous CH$_4$ and O$_2$/O$_3$ detection it is unnecessary to \textit{also} perform LUVOIR-B, or any architectures with smaller apertures, simulations.}
\end{enumerate}
\section{Results}\label{sec:result}
\remove{This work studied whether or not incoming water and oxygen fluxes into the TOA of a terrestrial planet could trigger a false-positive detection in transmission for upcoming and next-generation observatories. Here we present our results beginning with the photochemical mixing ratio and column densities and then the incoming water and the incoming oxygen flux synthetic spectra results. All PSG simulations were performed in cloudy or clear sky conditions and this is highlighted where appropriate.}
\subsection{\textit{Atmos} Results}\label{atmos}
\begin{figure}[H]
\centering
    \includegraphics[trim = 0 71 0 80, clip, width=\textwidth]{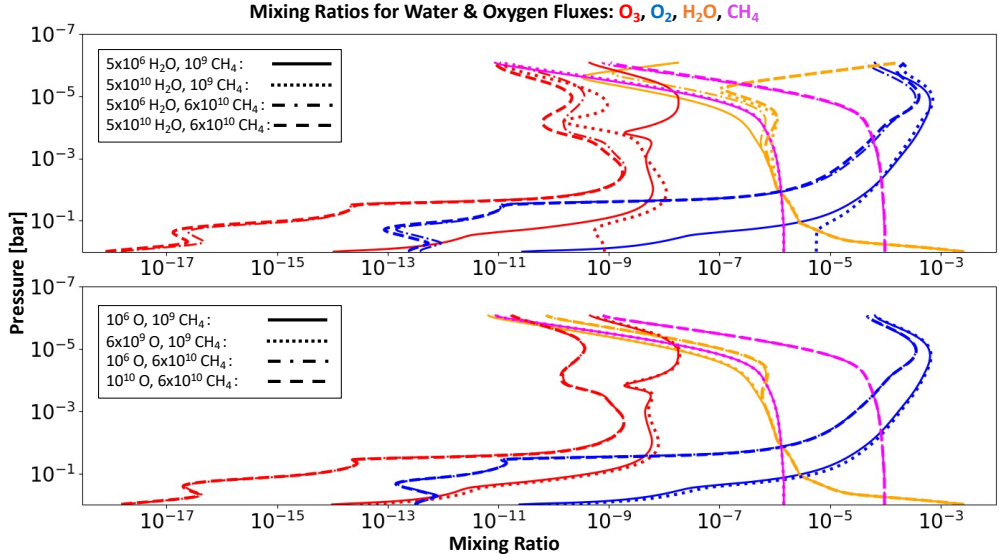}
    \caption{Mixing ratios for O$_3$, O$_2$\change{ and}{,} H$_2$O and CH$_4$ for the lowest and highest \textbf{water} fluxes and \textbf{oxygen} fluxes.}
    \label{fig:mixingratios}   
\end{figure}

\begin{figure}[H]
\centering
    \includegraphics[trim = 0 95 0 80, clip,width=\textwidth]{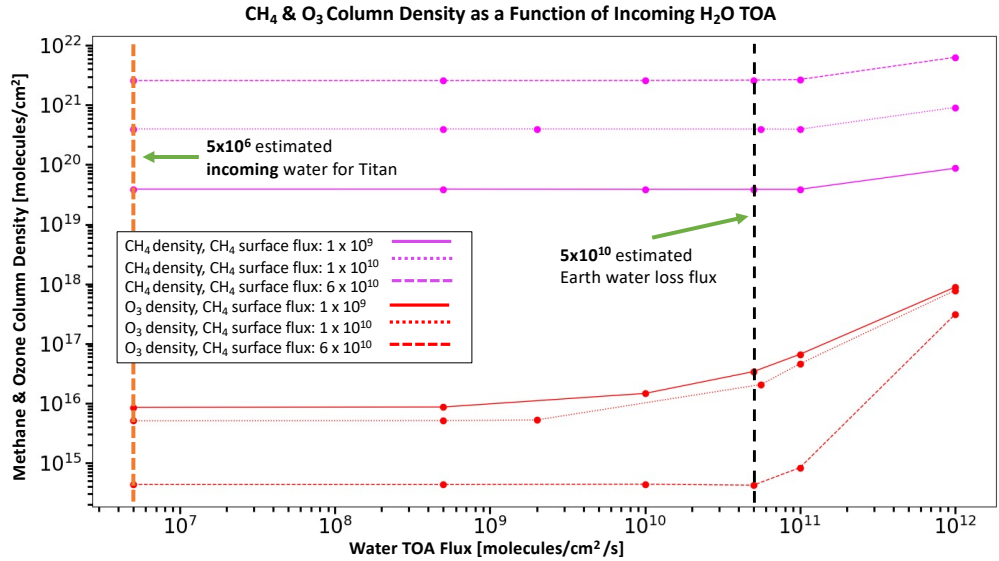}
    \caption{CH$_4$ and O$_3$ column density responses to increasing \textbf{water} flux values into the top of the atmosphere. The vertical dashed lines represent the lower and upper plausible limits of the water fluxes. The points to the right of the black vertical dashed line fall into the realm of physically implausible TOA fluxes. \add{All data points use the average of the temperature profiles. Note the lack of data points for the red and purple dotted lines at water flux values 10$^{10}$ and 5 $\times$ 10$^{10}$. These are regions where the photochemical model is not able to converge.}}
    \label{fig:cdensityvswater}
\end{figure}

\begin{figure}[H]
\centering
    \includegraphics[trim = 0 95 0 80, clip,width=\textwidth]{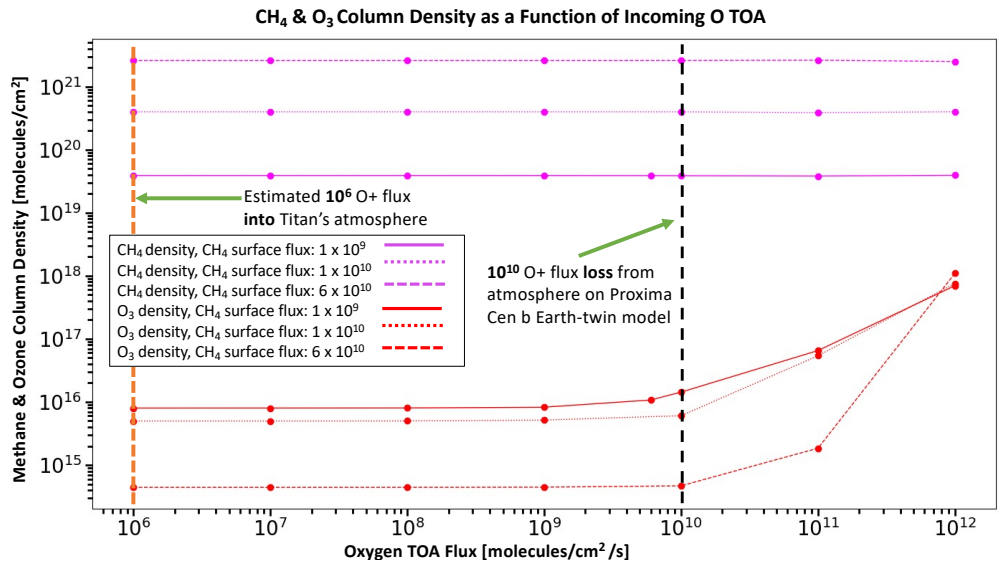}
    \caption{CH$_4$ and O$_3$ column density responses to increasing \textbf{oxygen} flux values into the top of the atmosphere. The vertical dashed lines represent the lower and upper limit of the oxygen fluxes.}
    \label{fig:cdensityvsoxygen}
\end{figure}

Our \add{coupled} photochemical\add{-climate} model results appear in Fig. \ref{fig:mixingratios}, \ref{fig:cdensityvswater}, and \ref{fig:cdensityvsoxygen}. In Fig. \ref{fig:mixingratios} we display our mixing ratios for O$_3$, O$_2$, H$_2$O and CH$_4$\remove{while varying the amount of O, H$_2$O and CH$_4$ that is entering the atmosphere from the top and from the surface}. \remove{While the }H$_2$O flux\add{es} (top panel) \change{has}{have} a more dramatic effect on the abundances of O$_3$, O$_2$, H$_2$O and CH$_4$, \add{than} the O fluxes (bottom panel). In the top panel a build up of H$_2$O in the upper layers of the atmosphere occurs, which is directly tied to the incoming TOA H$_2$O flux rising. Despite the large amount of incoming H$_2$O there was no super saturation in the top region of the atmosphere. When the CH$_4$ flux is low, increasing the water flux results in a higher percentage of O$_3$ and O$_2$ near the surface of the planet; when the CH$_4$ is high, varying the H$_2$O flux has little effect on the mixing ratios. \add{This divergence between O$_2$ and O$_3$ surface mixing ratios, when water flux increases, is due to catalytic cycles between OH, HO$_2$ and H$_2$O$_2$ and increasing escape of H and H$_2$ from the atmosphere. The escaping H and H$_2$ diminish both the reducing power of the atmosphere and the strength of the catalytic cycle, thus removing O$_2$ sinks. As the water flux increases and the atmosphere is unable to maintain the catalytic cycle, O$_2$ builds up in the lower half of the atmosphere. This will only occur when the atmosphere crosses a redox threshold where it is becoming oxidized with time, which in the case of our simulations occurs when the water or oxygen TOA flux is greater than 1 $\times$ 10$^{10}$ molecules/cm$^2$/s, which is why the behavior is not seen in the bottom panel of Fig. \ref{fig:mixingratios}.} In the bottom panel the only noticeable changes occur when the CH$_4$ flux is varied, while the differences between low and high O flux are insignificant.

\remove{In Fig. \ref{fig:cdensityvswater} and \ref{fig:cdensityvsoxygen}}\add{To show the effects of H$_2$O and O fluxes on the whole atmosphere,} we present\remove{our} CH$_4$ and O$_3$ column \remove{density results when H$_2$O or O is flowing in from the TOA}\add{densities as a function of H$_2$O flux (Fig. \ref{fig:cdensityvswater}) and O flux (Fig. \ref{fig:cdensityvsoxygen})}.\remove{The vertical dashed lines represent the end-member limits we chose for each flux.} \change{As the incoming water approaches the upper limit there is an approximately 4 order of magnitude jump in the O$_3$ column density as the H$_2$O flux goes from 1 $\times$ 10$^{10}$ to 5 $\times$ 10$^{10}$. Up until that point the O$_3$ column densities were all on the order of 10$^{12}$}{As the incoming water and oxygen approaches the upper limit, the simulations show a gradual rise in O$_3$ densities while CH$_4$ densities remain completely static; the densities are not reacting strongly to the rising fluxes}. The region on the far right of Fig. \ref{fig:cdensityvswater}, and \ref{fig:cdensityvsoxygen} shows the results when the physically plausible limits for water and oxygen fluxes are exceeded. It is important to note that these points past the limits are \textbf{not} included in the mixing ratio plots of Fig. \ref{fig:mixingratios}. Only when this limit is breached do\remove{es} the \add{simulated }atmosphere\add{s} begin to respond \add{more sharply} to the incoming oxygen \add{and water.} \add{This is}\remove{as} evident in the \change{abrupt 4 and 5 order of magnitude column density change in O$_3$ for the two lowest CH$_4$ surface fluxes (solid red and dotted red lines of Fig. \ref{fig:cdensityvsoxygen})}{1 to 3 order of magnitude changes in O$_3$ column densities for both water and oxygen fluxes. Conversely, the CH$_4$ is still responding slowly to the increasing fluxes (Fig. \ref{fig:cdensityvswater}) or remaining unperturbed (Fig. \ref{fig:cdensityvsoxygen})}\remove{While O$_3$ and CH$_4$ began to show noticeable column density changes as the water flux approaches 5 $\times$ 10$^{10}$, these species and the atmosphere as a whole are less sensitive to the increasing oxygen flowing into the TOA}.

Since our simulated atmospheres are receiving additional amounts of reducing and oxidizing material it is important to track the global ocean-atmosphere redox imbalance budgets to confirm redox is being conserved (Table \ref{redoxwater} and Table \ref{redoxoxygen}). \add{The redox imbalance represents the net reductant flux going into the atmosphere (+) or going into the ocean (-).} We follow the methodology described in \citeA{harman2018abiotic} and our results show the global ocean-atmosphere redox for the boundaries of our parameter space and the flux combinations that are used in PSG. Our photochemical model is able to simulate modern day Earth conditions, an atmosphere with large biological methane and oxygen fluxes. We compare these modern day Earth ocean-atmosphere redox results to our influx simulations as a check for consistency.

\begin{table}[H]
\caption{Coupled Model Global Ocean-Atmosphere Redox Imbalance Results - \textit{Water} Flux}
\centering
\begin{tabular}{l c c}
\hline
 H$_2$O flux & CH$_4$ flux & Ocean-Atmosphere Redox Imbalance\\
\hline
5 $\times$ 10$^{6}$ & 1 $\times$ 10$^{9}$ & -9.48$\times$ 10$^{9}$\\ 
5 $\times$ 10$^{10}$ & 1 $\times$ 10$^{9}$ & 7.66$\times$ 10$^{9}$\\ 
5 $\times$ 10$^{6}$ & 6 $\times$ 10$^{10}$ & 4.61$\times$ 10$^{9}$\\ 
5 $\times$ 10$^{10}$ & 6 $\times$ 10$^{10}$ & 2.24$\times$ 10$^{10}$\\ 
\hline
Modern Earth & 1 $\times$ 10$^{11}$ & -2.95$\times$ 10$^{8}$\\ 
\hline
\multicolumn{3}{p{0.9\columnwidth}}{\textbf{Note}. \add{Global ocean-atmosphere redox values for the photochemical simulations that are coupled \cite{harman2018abiotic}. The global ocean-atmosphere redox imbalance for the Modern Earth run with our photochemical model is also included as a reference for a world with large biological methane and oxygen fluxes.}}
\end{tabular}
\label{redoxwater}
\end{table}

\begin{table}[H]
\caption{Coupled Model Global Ocean-Atmosphere Redox Imbalance Results - \textit{Oxygen} Flux}
\centering
\begin{tabular}{l c c}
\hline
 O flux & CH$_4$ flux & Ocean-Atmosphere Redox Imbalance\\
\hline
1 $\times$ 10$^{6}$ & 1 $\times$ 10$^{9}$ & -9.47$\times$ 10$^{9}$\\ 
6 $\times$ 10$^{9}$ & 1 $\times$ 10$^{9}$ & -3.56$\times$ 10$^{9}$\\ 
1 $\times$ 10$^{6}$ & 6 $\times$ 10$^{10}$ & 4.63$\times$ 10$^{9}$\\ 
1 $\times$ 10$^{10}$ & 6 $\times$ 10$^{10}$ & 1.463$\times$ 10$^{10}$\\ 
\hline
Modern Earth & 1 $\times$ 10$^{11}$ & -2.95$\times$ 10$^{8}$\\ 
\hline
\multicolumn{3}{p{0.9\columnwidth}}{\textbf{Note}. \add{Coupled global ocean-atmosphere redox values when an oxygen flux into the TOA is present.}}
\end{tabular}
\label{redoxoxygen}
\end{table}

\subsection{Synthetic Spectra Features - CH$_4$, O$_3$, O$_2$}\label{spectra_results}
\begin{figure}[H]
\centering
   \includegraphics[trim = 0 90 0 75, clip,width=\textwidth]{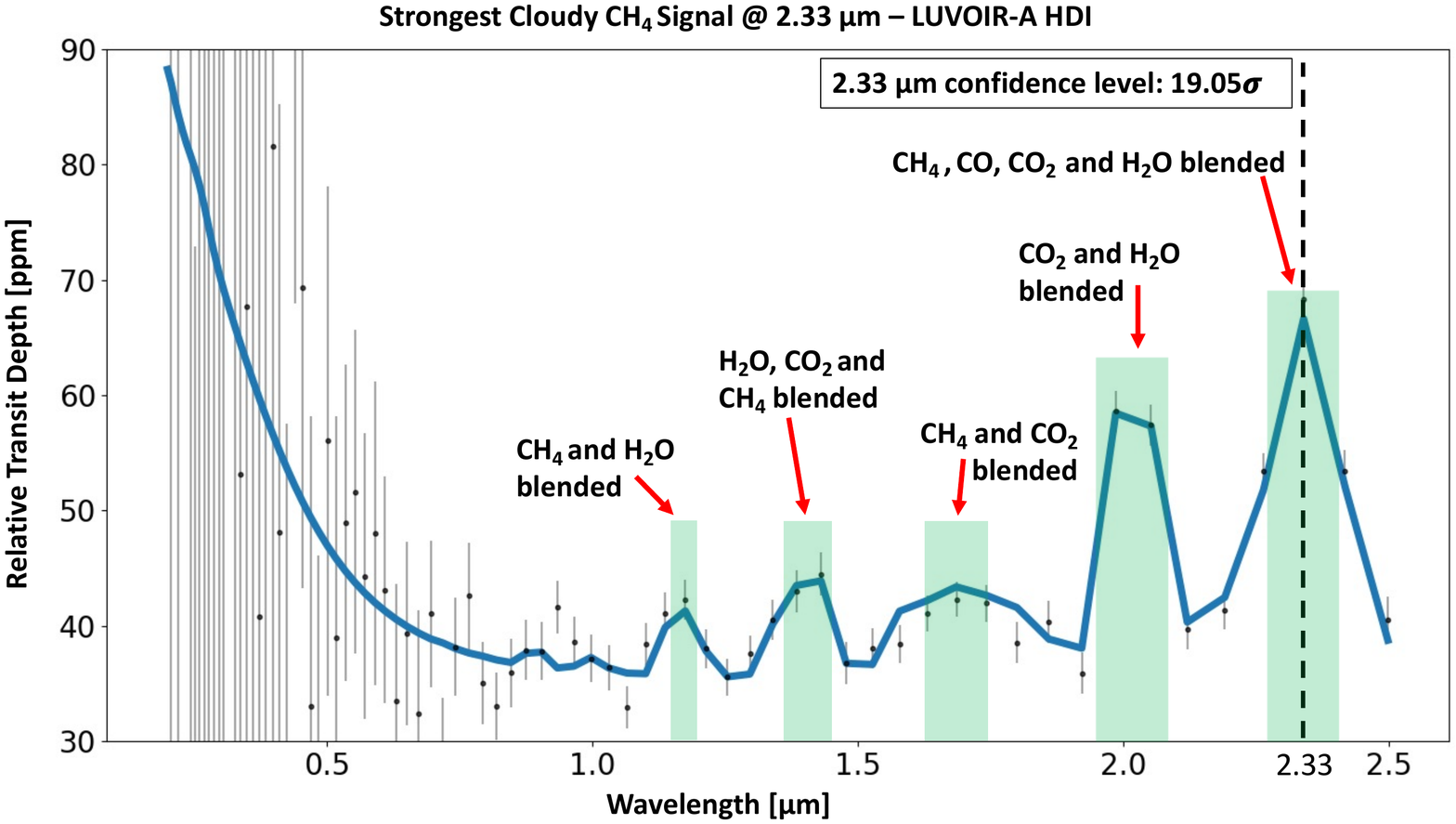}
   \includegraphics[trim = 0 90 0 75, clip,width=\textwidth]{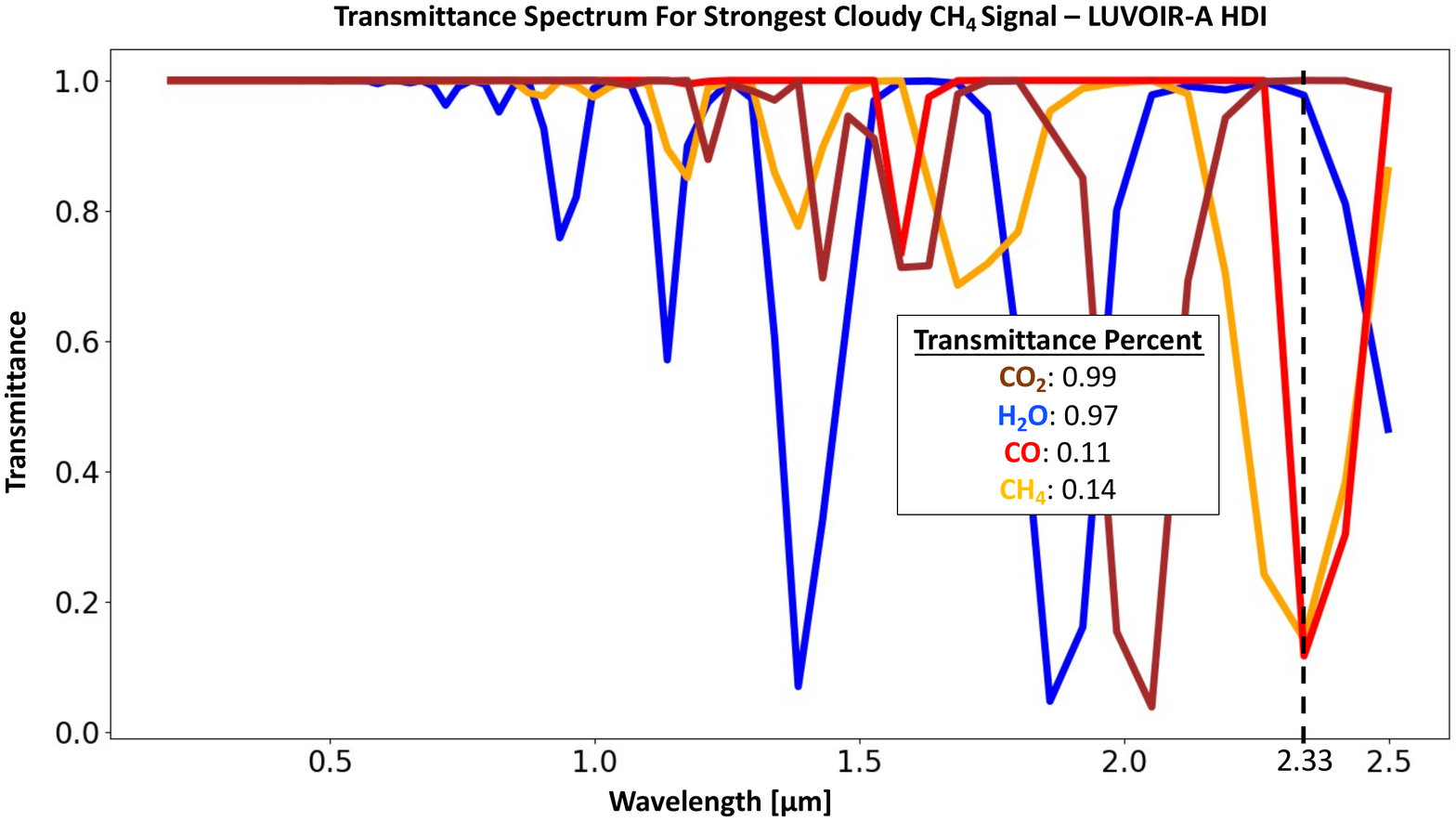}
        \caption{\remove{The transmission spectrum for the strongest and most reliable CH$_4$ feature produced for 85 transits, observed by LUVOIR-A HDI. Even though the 2.33 $\mu$m feature is blended with other gases, the CH$_4$ is still the most significant absorber at this wavelength. This is further evidence of a reliable and strong feature.}\textit{Top Panel}: Synthetic relative transit depth spectra of an atmosphere with an O TOA flux of 1 $\times$ 10$^{10}$ molecules/cm$^2$/s and a CH$_4$ surface flux of 6 $\times$ 10$^{10}$ molecules/cm$^2$/s. A \change{strong}{potentially detectable} CH$_4$ signal is \change{predicted}{simulated} at 2.33 $\mu$m. \textit{Bottom Panel}: The corresponding model transmission spectrum with the top four absorbers at 2.33 $\mu$m highlighted.}
        \label{luvoirch4hhcloudoxygen}
\end{figure}

The first PSG-calculated molecule-detection signal/noise results appear in Table \ref{alldetectability}. This shows the \change{signal-to-noise ratio (S/N)}{S/N} for all four observatories and their instruments under \add{both,} clear sky and cloudy conditions. We use the S/N as a metric to estimate the detectability of a given feature with a S/N of 5 being considered as reliable. \add{And S/N is shown for 5, 10 and 20 year mission lifetimes. Due to the speculative nature of mission lifetimes beyond $\sim$5 years we only highlight the 5 year results here within the text as the baseline.} The \change{strongest signal}{largest S/N} for \change{a realistic}{an} atmosphere with clouds \change{was}{is} produced for a LUVOIR-A HDI synthetic observation \add{(19.05$\sigma$)} \change{. The}{and the} relative transit depth (contrast) and transmittance spectra for this signal are shown in Fig. \ref{luvoirch4hhcloudoxygen}. This \change{was}{is} a CH$_4$ feature at 2.33 $\mu$m where \change{clouds did not play a large role}{the cloud opacity is weak} in the transmittance and a TOA \change{water}{oxygen} flux \change{was}{is} present. Under clear sky conditions (Table \ref{alldetectability}), \add{we predict that the following telescopes would detect a CH$_4$ feature due to a S/N $\geq$ 5: }Origins\remove{would be able to detect a CH$_4$ feature} at \change{3.3}{3.29} $\mu$m\add{, JWST NIRSpec PRISM at 3.37 $\mu$m and LUVOIR at 2.33 $\mu$m}\remove{while LUVOIR and JWST NIRSpec PRISM detect CH$_4$ at 2.33 $\mu$m}\remove{, all with at least a 5$\sigma$ predicted detection}.\remove{JWST MIRI-LRS showed a CH$_4$ feature at 7.72 $\mu$m and HabEx showed one at 1.72 $\mu$m but both were under 5$\sigma$and deemed to not be a very confident detection.} Under cloudy conditions \remove{(Fig. \ref{ch4clearcloud})}only LUVOIR\remove{and JWST NIRSpec PRISM} \change{detected a}{would detect a simulated} CH$_4$ feature (2.33 $\mu$m) \change{detectable with }{due to} a confidence level of at least 5$\sigma$. There were no instances of a \change{potential}{simulated} O$_3$ or O$_2$ feature\remove{being} detectable simultaneously with CH$_4$ within the physical limit bounds we established. \add{Our O$_2$ sensitivity scaling test also returned null results for O$_2$ or O$_3$ S/N. The only observatory that showed a simulated feature was LUVOIR at 0.25 $\mu$m but the noise was extremely high ($>$500 ppm) resulting in a S/N less than 1.} \add{The extended mission lifetime scaling resulted in additional simulated CH$_4$ features with a S/N $\geq$ 5 but did not produce any reliable O$_2$ or O$_3$ features (Table. \ref{alldetectability}).}

\remove{Even though the O$_2$ collision-induced signal at 6.4 $\mu$m has been proposed as one of the most promising O$_2$ features for transit observation \cite{fauchez2020sensitive} we did not detect it in this study. This is due to the nature of collision induced absorption (CIA) features and the low abundance of oxygen in our atmosphere. CIA is dependent on the square of the gas pressure so if we are in a region of the atmosphere with low pressure and very low oxygen abundance there will not be enough collisions to contribute to a signal.}

\begin{table}[H]
\caption{CH$_4$, O$_2$, O$_3$ Detectability Results}
\centering
\begin{tabular}{l c c c c}
\hline
& & & Cloudy & Clear Sky \\
Observatory & O flux & CH$_4$ flux & CH$_4$ S/N & CH$_4$ S/N \\
Potential Lifetimes [years] & & & 5, 10, 20 & 5, 10, 20 \\
\hline
  JWST MIRI-LRS & 1 $\times$ 10$^{10}$ & 6 $\times$ 10$^{10}$ & \change{$\sigma$ =}{2.19, 3.09, 3.20} & \change{$\sigma$ =}{3.93, \textbf{5.55}, \textbf{5.75}} \\
  JWST NIRSpec PRISM & 1 $\times$ 10$^{10}$ & 6 $\times$ 10$^{10}$ & \change{$\sigma$ =4.44}{2.34*} & \change{\textbf{$\sigma$ = 9.51}}{\textbf{5.02}*} \\  
  ORIGINS MISC-T & 1 $\times$ 10$^{10}$ & 6 $\times$ 10$^{10}$ & \change{$\sigma$ = }{4.36, \textbf{6.17}, \textbf{8.73}} & \change{$\sigma$ =}{\textbf{8.80}, \textbf{12.45}, \textbf{17.61}} \\
  HabEx HWC & 1 $\times$ 10$^{10}$ & 6 $\times$ 10$^{10}$ & \change{$\sigma$ = }{1.47, 2.08, 2.95} & \change{$\sigma$ = }{4.93, \textbf{6.97}, \textbf{9.86}} \\
  LUVOIR-A HDI & 1 $\times$ 10$^{10}$ & 6 $\times$ 10$^{10}$ & \textbf{19.05}, \textbf{26.94}, \textbf{38.11} & \textbf{34.79}, \textbf{49.21}, \textbf{69.59} \\
\hline
\hline
Observatory & H$_2$O flux & CH$_4$ flux & CH$_4$ S/N & CH$_4$ S/N\\
\hline
  JWST MIRI-LRS & 5 $\times$ 10$^{10}$ & 6 $\times$ 10$^{10}$ & \change{$\sigma$ = }{2.18, 3.09, 3.20} & \change{$\sigma$ = }{3.87, \textbf{5.47}, \textbf{5.67}}\\
  JWST NIRSpec PRISM & 5 $\times$ 10$^{10}$ & 6 $\times$ 10$^{10}$ & \change{$\sigma$ = }{2.35*} & \change{$\sigma$ = }{\textbf{5.03}*}  \\
  ORIGINS MISC-T & 5 $\times$ 10$^{10}$ & 6 $\times$ 10$^{10}$ & \change{$\sigma$ = }{4.38, \textbf{6.20}, \textbf{8.77}} & \change{$\sigma$ = }{\textbf{8.83}, \textbf{12.48}, \textbf{17.66}} \\
  HabEx HWC & 5 $\times$ 10$^{10}$ & 6 $\times$ 10$^{10}$ & \change{$\sigma$ = }{1.50, 2.13, 3.01} & \change{$\sigma$ = }{4.95, \textbf{7.00}, \textbf{9.91}} \\
  LUVOIR-A HDI & 5 $\times$ 10$^{10}$ & 6 $\times$ 10$^{10}$ & \textbf{18.01}, \textbf{25.47}, \textbf{36.02} & \textbf{33.75}, \textbf{47.73}, \textbf{67.50}\\  
\hline
\hline
 & O flux & CH$_4$ flux & O$_2$, O$_3$ S/N & O$_2$, O$_3$ S/N\\
\hline
 All Observatories & 6 $\times$ 10$^{9}$ & 1 $\times$ 10$^{9}$ & - & -\\
\hline
\hline
 & H$_2$O flux & CH$_4$ flux & O$_2$, O$_3$ S/N & O$_2$, O$_3$ S/N\\
\hline
 All Observatories & 5 $\times$ 10$^{10}$ & 1 $\times$ 10$^{9}$ & - & -\\ 
\hline
\multicolumn{5}{p{0.9\columnwidth}}{\textbf{Note}. S/N results for all observatories studied. \add{The S/N is computed for the \textit{in transit} time only, without considering the out of transit observation and the values are for 85, 170 and 340 transits.} The \textbf{bolded} values represent simulated detections with confidence levels of 5$\sigma$ or higher. All O$_3$ and O$_2$ features are predicted to be non-existent \change{or}{, blended too heavily with other gases or} completely enveloped by the synthetic noise. Flux is in units of molecules/cm$^2$/s. \add{*Asterisk's denote JWST S/N values that triggered the estimated 10ppm noise floor in the first 5 years and thus are unable to improve their S/N with longer mission lifetimes (more transits).}}
\end{tabular}
\label{alldetectability}
\end{table}

\subsection{Synthetic Spectra - Incoming Oxygen $\&$ Water Past the Limits}\label{sec:take_it_to_the_limit}
Even though no O$_3$\change{(O$_2$)}{ or O$_2$} synthetic signal \change{was}{is} produced within the bounds set in the column density plots from Fig. \ref{fig:cdensityvswater} and \ref{fig:cdensityvsoxygen}, \change{we were still curious as to what}{we estimate how much} O or H$_2$O flux is required to cause a potentially detectable O$_3$\change{(O$_2$)}{/O$_2$} signal. When the physical flux bounds were surpassed, simulated features appear within the wavelength range of JWST MIRI-LRS \change{LUVOIR-A and Origins}{and LUVOIR-A} but their transit depths and S/N are below the level of a reliable detection (Table \ref{implausibleflux}). For JWST MIRI-LRS \remove{and Origins}there is a simulated O$_3$ feature at $\sim$9.6 $\mu$m but the S/N is smaller than 5$\sigma$. Additionally, there is a 0.25 $\mu$m feature in LUVOIR-A's wavelength range; however, due to extreme noise in the UV the resulting S/N for this feature is less then 1$\sigma$. \add{When the extended mission lifetimes are taken into account and the physically plausible flux ranges are breached, Origins MISC-T is able to simulate an O$_3$ feature at 9.73 $\mu$m with an S/N $>$ 5 after 10 years (170 transits).}

\begin{table}[H]
\caption{Detections in Physically Implausible Flux Range}
\centering
\begin{tabular}{l c c c c}
\hline
& & & Cloudy & Clear Sky \\
Observatory & O flux & CH$_4$ flux & O$_3$ S/N & O$_3$ S/N\\
Potential Lifetimes [years] & & & 5, 10, 20 & 5, 10, 20 \\
\hline
\hline
JWST MIRI-LRS & 1 $\times$ 10$^{12}$ & 1 $\times$ 10$^{9}$ & 0.65, 0.92, 1.24 & 1.90, 2.69, 3.62 \\
  JWST NIRSpec PRISM & 1 $\times$ 10$^{12}$ & 1 $\times$ 10$^{9}$ & - &  - \\
  ORIGINS MISC-T & 1 $\times$ 10$^{12}$ & 1 $\times$ 10$^{9}$ & 1.00, 1.42, 2.01 & 3.03, 4.29, 6.07 \\
  HabEx HWC & 1 $\times$ 10$^{12}$ & 1 $\times$ 10$^{9}$ & - & - \\
  LUVOIR-A HDI & 1 $\times$ 10$^{12}$ & 1 $\times$ 10$^{9}$ & 0.09, 0.14, 0.19 & 0.14, 0.20, 0.29 \\ 
\hline
\hline
Observatory & H$_2$O flux & CH$_4$ flux & O$_3$ S/N & O$_3$ S/N\\
JWST MIRI-LRS & 1 $\times$ 10$^{12}$ & 1 $\times$ 10$^{9}$ & 0.84, 1.19, 1.61 & 2.08, 2.94, 3.95 \\
JWST NIRSpec PRISM & 1 $\times$ 10$^{12}$ & 1 $\times$ 10$^{9}$ & - & - \\
ORIGINS MISC-T & 1 $\times$ 10$^{12}$ & 1 $\times$ 10$^{9}$ & 1.35, 1.91, 2.71 & 3.46, 4.90, 6.93\\
HabEx HWC & 1 $\times$ 10$^{12}$ & 1 $\times$ 10$^{9}$ & - & - \\
LUVOIR-A HDI & 1 $\times$ 10$^{12}$ & 1 $\times$ 10$^{9}$ & 0.09, 0.13, 0.19 & 0.14, 0.20, 0.28 \\  
\hline
\hline
\multicolumn{5}{p{0.9\columnwidth}}{\textbf{Note}. Synthetic simulations in the physically implausible parameter space from Fig \ref{fig:cdensityvswater} and \ref{fig:cdensityvsoxygen}. The CH$_4$ flux uses the abiotic value of 1 $\times$ 10$^9$ molecules/cm$^2$/s. Under these conditions, \change{there are no observatories that produce a reliable synthetic observation (S/N $\geq$ 5) of O$_2$ or O$_3$.}{If the physically plausible flux ranges are breached and a mission lifetime extends past 10 years, then Origins MISC-T produces a synthetic O$_3$ observation, under clear sky conditions, with an S/N $>$ 5.}}
\end{tabular}
\label{implausibleflux}
\end{table}

\subsection{Synthetic Spectra - Abiotic CH$_4$ Detections}\label{sec:how_low_can_you_go}
At the abiotic CH$_4$ flux level (1 $\times$ 10$^9$) we consider in our work, CH$_4$ features are present but all of them are at a S/N lower than 5$\sigma$ and/or heavily blended with other gases such as H$_2$O and CO \add{except when simulating ORIGINS MISC-T extended lifetime observations} (Table \ref{ch4lowest}). \add{Under clear sky conditions we simulate ORIGINS detecting a 3.29 $\mu$m CH$_4$ with a S/N larger than 5$\sigma$.} \remove{CH$_4$ is never the dominant feature in any of the blended cases and even for a large aperture observatory such as LUVOIR-A the results are not promising.}We \add{also }predict a simulated CH$_4$ feature in clear sky conditions for LUVOIR-A at 2.33 $\mu$m with a S/N over 21$\sigma$; however, the transmittance spectrum shows that this feature is predominately CO at 70$\%$. \add{No potentially reliable simulated feature was detected under cloudy conditions.}

\begin{table}[H] 
\caption{CH$_4$ S/N For Lowest CH$_4$, H$_2$O $\&$ O Fluxes}
\centering
\begin{tabular}{l c c c c}
\hline
& & & Cloudy & Clear Sky \\
 & O flux & CH$_4$ flux & CH$_4$ S/N & CH$_4$ S/N\\
Potential Lifetimes [years] & & & 5, 10, 20 & 5, 10, 20 \\ 
\hline
\hline
ORIGINS MISC-T & 1 $\times$ 10$^{6}$ & 1 $\times$ 10$^{9}$ & - & 4.16, \textbf{5.88}, \textbf{8.32}\\ 
\hline
 & H$_2$O flux & CH$_4$ flux & CH$_4$ S/N & CH$_4$ S/N\\
\hline
ORIGINS MISC-T & 5 $\times$ 10$^{6}$ & 1 $\times$ 10$^{9}$ & - & 4.16, \textbf{5.88}, \textbf{8.32}\\
\hline
\multicolumn{5}{p{0.9\columnwidth}}{\textbf{Note}. S/N CH$_4$ results when the CH$_4$ surface flux was 1 $\times$ 10$^9$ molecules/cm$^2$/s which is considered an abiotic flux. Using these fluxes, we predict that CH$_4$ \change{would not be detectable for any of the observatories studied}{is detectable under clear sky conditions in the extended lifetime scenario for ORIGINS MISC-T for both incoming flux types. \textbf{Bolded} values represent simulated detections with confidence levels of 5$\sigma$ or higher}.}
\end{tabular}
\label{ch4lowest}
\end{table}

\section{Discussion}\label{sec:disc}
To avoid misidentifying biosignatures it is important to be aware of the many false-positives that can arise given the complexity of terrestrial exoplanets that we expect to encounter. We explore\add{d} here whether the mechanism of volatiles flowing into a terrestrial planet's atmosphere could produce a false positive by raising the levels of oxygen and ozone to abundances that would be misinterpreted as originating from biology. Our results indicate that this mechanism would not produce a detectable false-positive biosignature when observed by JWST or a next-generation observatory. 

To simulate the incoming material flow of volatiles we needed to simulate a range of fluxes to the TOA. When doing so, we sought to set an aggressive value for the upper end of our flux range, to ensure our range of fluxes included all possible abiotic fluxes. We envisioned a scenario where a planet, such as TRAPPIST-1 d, orbiting an M dwarf was losing its atmosphere and that the next closet planet in the direction of the outflow was gaining it into their atmosphere, which in this case would be TRAPPIST-1 e. An M-dwarf system was chosen due to the ubiquitousness of M-dwarfs in the galaxy, the close-in region of the habitable zone around M-dwarfs and the tendency for increased stellar activity compared to G-stars which would lead to a higher chance of planets losing their atmospheres. Active M-dwarfs with their extreme ultraviolet output and flares are likely to increase atmospheric escape on any of their close-in planets causing the planets to lose their atmosphere. Even under such extreme assumptions, where every molecule outgassed from one planet ends up entering the second, \add{we did not predict a detectable O$_2$/O$_3$ signal}\remove{there was not a detectable O$_3$ signal}. \remove{A lack of an O$_3$ signal also led to there not being a detectable O$_2$ signal. This follows the work of \cite{segura2003ozone} and \cite{harman2015abiotic} which shows that O$_3$ becomes observable before O$_2$ even when the O$_3$ abundances are less than the O$_2$ abundances. Specifically Table 5 and Figure 9 from \cite{harman2015abiotic} show how the O$_2$ column depth needs to be roughly 5 orders of magnitude above the O$_3$ column depth for a signal to begin appearing. In our work the difference in column depths only approached 3 orders of magnitude and were not close to present day earth abundances, so it is ultimately not a surprise for us to only detect \textit{weak} O$_3$ features originating from physically implausible TOA fluxes.}

\subsection{Oxygen \& Water\change{Discussion}{'s Role in Detectability}}
The results allow for a strong conclusion: fluxes of O into a planetary atmosphere will not lead to a detectable O$_3$ signal in an atmosphere that also has appreciable CH$_4$. The current understanding of O$_3$ detectability is that in atmospheres with detectable CH$_4$, O$_3$ will not be detectable for abiotic O$_2$ fluxes, which are significantly smaller than the O$_2$ flux from modern day biology $\sim$4 $\times$ 10$^{13}$ molecules/cm$^2$/s \cite{walker1974stability}. By considering that prior work and our results here, there is a consistent conclusion. Whether O is flowing into the atmosphere at the bottom or the top, the rates required to have O$_3$ accumulate to detectable levels in a CH$_4$-rich atmosphere are \add{orders of magnitude }higher than can be achieved through abiotic processes. Thus, gas fluxes - if they can be constrained - remain a very strong biosignature. Similarly, \remove{the upper limit on}\add{we predict} water influx\add{es} \change{was also not sufficient to}{will not} generate detectable O$_3$ concentrations. The incoming water flux does \add{produce potentially detectable O$_3$ when extended mission lifetimes are considered but \textit{only} at implausibly high fluxes.}\remove{increase the amount of O$_3$ at implausibly high fluxes, but only at undetectable O$_3$ levels.} Thus, we can confidently say that an incoming water flux will not lead to a false-positive for the CH$_4$-\change{O$_3$ (O$_2$)}{O$_2$/O$_3$} biosignature pair. \remove{Despite the plausibility of a water molecule surviving the trip from outer space and into an atmosphere wholly intact, we modelled the experiment this way for multiple reasons. Using water molecules as a representation for oxygen molecule flux into an atmosphere has been used by Titan photochemical modelers \cite{hebrard2012neutral}. There is also the stoichiometric argument to be made that even if all of the water molecules were dissociated on their way to the top of a planet's atmosphere, the hydrogen atoms should still be present. Perhaps most importantly, by modelling a water flux we are able to account for other external oxygen sources to an atmosphere such as water ice.}

\subsection{Methane Detectability, Sensitivity Scaling of O$_2$ $\&$ Next Steps, JWST Scaling }
Although this was not the main motivation for this study, we were also able to consider the impact of TOA H$_2$O and O fluxes on the detectability of CH$_4$. As Table \ref{ch4lowest} shows, we were \change{unable}{able} to predict the detection of a \add{potentially} reliable synthetic abiotic CH$_4$ feature \add{when extended mission lifetimes were considered}. \add{Assuming an ORIGINS-like Probe-class mission is developed and launched, as the decadal survey recommends, then we can expect it to be able to detect abiotic amounts of CH$_4$ in a dry abiotic terrestrial atmosphere.} Though LUVOIR-A saw a signal at 2.33 $\mu$m under clear sky conditions and the S/N was larger than 5$\sigma$, the issue becomes separating the signal into the CH$_4$ and CO components which already heavily favor the CO based on the transmittance results. If a desiccated (i.e. clear sky conditions) Earth-sized world was being observed and CO could be constrained then a next-generation observatory on the scale of LUVOIR-A may \add{also} be able to detect abiotic amounts of CH$_4$. This \remove{also }leads to the conclusion that neither JWST or any of the potential next-generation observatories could reliably detect abiotic CH$_4$ in a \textit{wet} abiotic terrestrial atmosphere. Nevertheless, we urge caution in ascribing a biological source of CH$_4$ unless the CH$_4$ itself can be constrained to be at levels only consistent with a global biosphere.

\add{Even though TRAPPIST-1 e is likely to be a synchronous rotator, by using a 1D model we negated the potential day and night side atmosphere discrepancies that could arise on such a body. However, our sensitivity scaling test on the accumulation of O$_2$ showed no effect on the detectability of this species. In addition, \protect\citeA{chen2018biosignature, chen2019habitability}} \add{have shown that major atmospheric species of interest to our work (CH$_4$ and O$_3$) do not experience order of magnitude mixing ratio differences between the day and night side of tidally locked worlds orbiting active or quiescent M dwarfs.}

While this study focused on the ability of TOA fluxes to create a ``false positive" for life, additional studies \change{could}{may} be performed to see how observations could detect and constrain the TOA fluxes. Such studies should make predictions of the observable impacts of this planetary process. This would allow future observatories to put these processes, observed on Titan, into a comparative planetology framework. \add{The temperature profiles of our simulations could also be further improved with the coupling of a GCM. The inclusion of 1D climate coupling was a critical first step to simulating reasonable temperature profiles but the lack of cloud feedbacks on the temperature and atmosphere mean both the photochemical and spectral simulations can be further refined. Finally, we focused exclusively on space based observatories but next-generation ground based telescopes such as the European-Extremely Large Telescope (E-ELT) \cite{gilmozzi2007european}, Giant Magellan Telescope (GMT) \cite{johns2006giant} and the Thirty Meter Telescope (TMT) \cite{skidmore2015thirty} may also have a role to play in Earth-sized exoplanet characterization endeavors. This is highly dependent on the outcome of technological breakthroughs that are being tested at this time. For example, the E-ELT was recently upgraded as part of the New Earths in the Alpha Centauri Region (NEAR) campaign through the Breakthrough Initiatives. This upgrade consisted of experimental starlight suppression techniques while observing in the mid-infrared and the results currently stand at a potential minimum detection range of $\sim$3 Earth-radii \cite{carlomagno2020metis,wagner2021imaging}. If this technology trend continues and the upcoming class of extremely large ground telescopes can reach smaller radii capabilities, then simulated false-positive biosignature detection comparisons between ground and space-based observatories would be an appropriate next step.} \remove{The photochemical modelling could also be improved with climate coupling that would allow the temperature and pressure to change in response to the atmosphere.} \remove{The temperature could also be improved with the use of a GCM that includes clouds since the inclusion or lack of clouds would affect the temperature profiles.}

\section{Conclusions}\label{sec:conclusion}
Our study shows that for transmission spectroscopy observations of terrestrial exoplanets orbiting M-dwarfs, \change{material exchange of water or oxygen between planets}{the transfer of water or oxygen from one planet to another} will not trigger a \add{detectable} false-positive for the biosignature consisting of the simultaneous presence of CH$_4$ and \change{O$_3$ (O$_2$)}{O$_2$/O$_3$} in a planetary atmosphere. The fluxes (1 $\times$ 10$^{12}$ molecules/cm$^2$/s) required to do so would need to be \change{at least}{more than} two orders of magnitude larger than what abiotic processes would allow.

\acknowledgments
DATA AVAILABILITY: All photochemical model runs, input files, and PSG configuration files are accessible at the public repository: \remove{https://doi.org/10.5281/zenodo.5591912}\add{\protect\citeA{feltonzenodowater2021}.}

Goddard affiliates acknowledge support from the GSFC Sellers Exoplanet Environments Collaboration (SEEC), which is funded in part by the NASA Planetary Science Division’s Internal Scientist Funding Model. This work was performed as part of NASA’s Virtual Planetary Laboratory, supported by the National Aeronautics and Space Administration through the NASA Astrobiology Institute under solicitation NNH12ZDA002C and Cooperative Agreement Number NNA13AA93A, and by the NASA Astrobiology Program under grant 80NSSC18K0829 as part of the Nexus for Exoplanet System Science (NExSS) research coordination network.\add{ RCF and STB acknowledge support by NASA under award number 80GSFC21M0002.} KEM and ALK acknowledge support from NASA NFDAP 80NSSC18K1233.

\remove{This work made use of the Python coronagraph noise model, developed by J. Lustig-Yaeger and available at https://github.com/jlustigy/coronagraph/.}

\textbf{RCF} performed all photochemical and PSG model runs, led analysis of the results and drafted the manuscript. \textbf{SDDG} helped conceptualize the project, and with \textbf{STB} provided guidance on the photochemical simulations and interpretations thereof, and provided feedback on the exoplanet and observatory implications of the work. \add{\textbf{STB} also performed all coupled photochemistry-climate simulations.} \textbf{KEM} and \textbf{ALK} provided insights into the transport of oxygen and water molecules in the Saturn system as well as guidance on determining reasonable upper limits for escape of heavy molecules from atmospheres. \textbf{TJF} provided inputs on the simulation of the transmission spectra and on the detectability of molecular signatures as well as on the observatories. \textbf{Everyone} contributed to the edits and comments of the entire manuscript.

\add{We would like to thank Jaime Crouse for the photolysis cross-section updates she made to the  photochemical-climate model. }We would also like to acknowledge the following individuals for their insights and guidance: Giada Arney, Chester ``Sonny" Harman, Marc Neveu, Daria Pidhorodetska, Sean Terry and Geronimo Villanueva.

\bibliography{water}

\begin{thebibliography}{}

\bibitem [\protect \citeauthoryear {%
Agol%
\ \protect \BOthers {.}}{%
Agol%
\ \protect \BOthers {.}}{%
{\protect \APACyear {2021}}%
}]{%
agol2021refining}
\APACinsertmetastar {%
agol2021refining}%
\begin{APACrefauthors}%
Agol, E.%
, Dorn, C.%
, Grimm, S\BPBI L.%
, Turbet, M.%
, Ducrot, E.%
, Delrez, L.%
\BDBL {}others%
\end{APACrefauthors}%
\unskip\
\newblock
\APACrefYearMonthDay{2021}{}{}.
\newblock
{\BBOQ}\APACrefatitle {{Refining the transit-timing and photometric analysis of
  TRAPPIST-1: masses, radii, densities, dynamics, and ephemerides}} {{Refining
  the transit-timing and photometric analysis of TRAPPIST-1: masses, radii,
  densities, dynamics, and ephemerides}}.{\BBCQ}
\newblock
\APACjournalVolNumPages{The planetary science journal}{2}{1}{1}.
\PrintBackRefs{\CurrentBib}

\bibitem [\protect \citeauthoryear {%
Arney%
, Domagal-Goldman%
\BCBL {}\ \BBA {} Meadows%
}{%
Arney%
\ \protect \BOthers {.}}{%
{\protect \APACyear {2018}}%
}]{%
arney2018organic}
\APACinsertmetastar {%
arney2018organic}%
\begin{APACrefauthors}%
Arney, G.%
, Domagal-Goldman, S\BPBI D.%
\BCBL {}\ \BBA {} Meadows, V\BPBI S.%
\end{APACrefauthors}%
\unskip\
\newblock
\APACrefYearMonthDay{2018}{}{}.
\newblock
{\BBOQ}\APACrefatitle {{Organic haze as a biosignature in anoxic Earth-like
  atmospheres}} {{Organic haze as a biosignature in anoxic Earth-like
  atmospheres}}.{\BBCQ}
\newblock
\APACjournalVolNumPages{Astrobiology}{18}{3}{311--329}.
\PrintBackRefs{\CurrentBib}

\bibitem [\protect \citeauthoryear {%
Arney%
\ \protect \BOthers {.}}{%
Arney%
\ \protect \BOthers {.}}{%
{\protect \APACyear {2016}}%
}]{%
arney2016pale}
\APACinsertmetastar {%
arney2016pale}%
\begin{APACrefauthors}%
Arney, G.%
, Domagal-Goldman, S\BPBI D.%
, Meadows, V\BPBI S.%
, Wolf, E\BPBI T.%
, Schwieterman, E.%
, Charnay, B.%
\BDBL {}Trainer, M\BPBI G.%
\end{APACrefauthors}%
\unskip\
\newblock
\APACrefYearMonthDay{2016}{}{}.
\newblock
{\BBOQ}\APACrefatitle {The pale orange dot: the spectrum and habitability of
  hazy Archean Earth} {The pale orange dot: the spectrum and habitability of
  hazy archean earth}.{\BBCQ}
\newblock
\APACjournalVolNumPages{Astrobiology}{16}{11}{873--899}.
\PrintBackRefs{\CurrentBib}

\bibitem [\protect \citeauthoryear {%
Arney%
\ \protect \BOthers {.}}{%
Arney%
\ \protect \BOthers {.}}{%
{\protect \APACyear {2017}}%
}]{%
arney2017pale}
\APACinsertmetastar {%
arney2017pale}%
\begin{APACrefauthors}%
Arney, G.%
, Meadows, V\BPBI S.%
, Domagal-Goldman, S\BPBI D.%
, Deming, D.%
, Robinson, T\BPBI D.%
, Tovar, G.%
\BDBL {}Schwieterman, E.%
\end{APACrefauthors}%
\unskip\
\newblock
\APACrefYearMonthDay{2017}{}{}.
\newblock
{\BBOQ}\APACrefatitle {Pale orange dots: the impact of organic haze on the
  habitability and detectability of earthlike exoplanets} {Pale orange dots:
  the impact of organic haze on the habitability and detectability of earthlike
  exoplanets}.{\BBCQ}
\newblock
\APACjournalVolNumPages{ApJ}{836}{1}{49}.
\PrintBackRefs{\CurrentBib}

\bibitem [\protect \citeauthoryear {%
Badhan%
\ \protect \BOthers {.}}{%
Badhan%
\ \protect \BOthers {.}}{%
{\protect \APACyear {2019}}%
}]{%
badhan2019stellar}
\APACinsertmetastar {%
badhan2019stellar}%
\begin{APACrefauthors}%
Badhan, M\BPBI A.%
, Wolf, E\BPBI T.%
, Kopparapu, R\BPBI K.%
, Arney, G.%
, Kempton, E\BPBI M\BHBI R.%
, Deming, D.%
\BCBL {}\ \BBA {} Domagal-Goldman, S\BPBI D.%
\end{APACrefauthors}%
\unskip\
\newblock
\APACrefYearMonthDay{2019}{}{}.
\newblock
{\BBOQ}\APACrefatitle {Stellar Activity Effects on Moist Habitable Terrestrial
  Atmospheres around M Dwarfs} {Stellar activity effects on moist habitable
  terrestrial atmospheres around m dwarfs}.{\BBCQ}
\newblock
\APACjournalVolNumPages{The Astrophysical Journal}{887}{1}{34}.
\PrintBackRefs{\CurrentBib}

\bibitem [\protect \citeauthoryear {%
Barnes%
}{%
Barnes%
}{%
{\protect \APACyear {2017}}%
}]{%
barnes2017tidal}
\APACinsertmetastar {%
barnes2017tidal}%
\begin{APACrefauthors}%
Barnes, R.%
\end{APACrefauthors}%
\unskip\
\newblock
\APACrefYearMonthDay{2017}{}{}.
\newblock
{\BBOQ}\APACrefatitle {Tidal locking of habitable exoplanets} {Tidal locking of
  habitable exoplanets}.{\BBCQ}
\newblock
\APACjournalVolNumPages{Celestial mechanics and dynamical
  astronomy}{129}{4}{509--536}.
\PrintBackRefs{\CurrentBib}

\bibitem [\protect \citeauthoryear {%
Berndt%
, Allen%
\BCBL {}\ \BBA {} Seyfried~Jr%
}{%
Berndt%
\ \protect \BOthers {.}}{%
{\protect \APACyear {1996}}%
}]{%
berndt1996reduction}
\APACinsertmetastar {%
berndt1996reduction}%
\begin{APACrefauthors}%
Berndt, M\BPBI E.%
, Allen, D\BPBI E.%
\BCBL {}\ \BBA {} Seyfried~Jr, W\BPBI E.%
\end{APACrefauthors}%
\unskip\
\newblock
\APACrefYearMonthDay{1996}{}{}.
\newblock
{\BBOQ}\APACrefatitle {Reduction of CO2 during serpentinization of olivine at
  300 C and 500 bar} {Reduction of co2 during serpentinization of olivine at
  300 c and 500 bar}.{\BBCQ}
\newblock
\APACjournalVolNumPages{Geology}{24}{4}{351--354}.
\PrintBackRefs{\CurrentBib}

\bibitem [\protect \citeauthoryear {%
Brown%
, Lebreton%
\BCBL {}\ \BBA {} Waite%
}{%
Brown%
\ \protect \BOthers {.}}{%
{\protect \APACyear {2009}}%
}]{%
brown2009titan}
\APACinsertmetastar {%
brown2009titan}%
\begin{APACrefauthors}%
Brown, R.%
, Lebreton, J\BPBI P.%
\BCBL {}\ \BBA {} Waite, H.%
\end{APACrefauthors}%
\unskip\
\newblock
\APACrefYear{2009}.
\newblock
\APACrefbtitle {Titan from Cassini-Huygens} {Titan from cassini-huygens}.
\newblock
\APACaddressPublisher{}{Springer Science \& Business Media}.
\PrintBackRefs{\CurrentBib}

\bibitem [\protect \citeauthoryear {%
Carlomagno%
\ \protect \BOthers {.}}{%
Carlomagno%
\ \protect \BOthers {.}}{%
{\protect \APACyear {2020}}%
}]{%
carlomagno2020metis}
\APACinsertmetastar {%
carlomagno2020metis}%
\begin{APACrefauthors}%
Carlomagno, B.%
, Delacroix, C.%
, Absil, O.%
, Cantalloube, F.%
, de Xivry, G\BPBI O.%
, Pathak, P.%
\BDBL {}others%
\end{APACrefauthors}%
\unskip\
\newblock
\APACrefYearMonthDay{2020}{}{}.
\newblock
{\BBOQ}\APACrefatitle {METIS high-contrast imaging: design and expected
  performance} {Metis high-contrast imaging: design and expected
  performance}.{\BBCQ}
\newblock
\APACjournalVolNumPages{Journal of Astronomical Telescopes, Instruments, and
  Systems}{6}{3}{035005}.
\PrintBackRefs{\CurrentBib}

\bibitem [\protect \citeauthoryear {%
Chen%
, Wolf%
, Kopparapu%
, Domagal-Goldman%
\BCBL {}\ \BBA {} Horton%
}{%
Chen%
\ \protect \BOthers {.}}{%
{\protect \APACyear {2018}}%
}]{%
chen2018biosignature}
\APACinsertmetastar {%
chen2018biosignature}%
\begin{APACrefauthors}%
Chen, H.%
, Wolf, E\BPBI T.%
, Kopparapu, R.%
, Domagal-Goldman, S.%
\BCBL {}\ \BBA {} Horton, D\BPBI E.%
\end{APACrefauthors}%
\unskip\
\newblock
\APACrefYearMonthDay{2018}{}{}.
\newblock
{\BBOQ}\APACrefatitle {Biosignature anisotropy modeled on temperate tidally
  locked m-dwarf planets} {Biosignature anisotropy modeled on temperate tidally
  locked m-dwarf planets}.{\BBCQ}
\newblock
\APACjournalVolNumPages{The Astrophysical Journal Letters}{868}{1}{L6}.
\PrintBackRefs{\CurrentBib}

\bibitem [\protect \citeauthoryear {%
Chen%
, Wolf%
, Zhan%
\BCBL {}\ \BBA {} Horton%
}{%
Chen%
\ \protect \BOthers {.}}{%
{\protect \APACyear {2019}}%
}]{%
chen2019habitability}
\APACinsertmetastar {%
chen2019habitability}%
\begin{APACrefauthors}%
Chen, H.%
, Wolf, E\BPBI T.%
, Zhan, Z.%
\BCBL {}\ \BBA {} Horton, D\BPBI E.%
\end{APACrefauthors}%
\unskip\
\newblock
\APACrefYearMonthDay{2019}{}{}.
\newblock
{\BBOQ}\APACrefatitle {Habitability and spectroscopic observability of warm
  M-dwarf exoplanets evaluated with a 3D chemistry-climate model} {Habitability
  and spectroscopic observability of warm m-dwarf exoplanets evaluated with a
  3d chemistry-climate model}.{\BBCQ}
\newblock
\APACjournalVolNumPages{The Astrophysical Journal}{886}{1}{16}.
\PrintBackRefs{\CurrentBib}

\bibitem [\protect \citeauthoryear {%
Cooray%
\ \protect \BOthers {.}}{%
Cooray%
\ \protect \BOthers {.}}{%
{\protect \APACyear {2019}}%
}]{%
cooray2019origins}
\APACinsertmetastar {%
cooray2019origins}%
\begin{APACrefauthors}%
Cooray, A.%
, Meixner, M.%
, Leisawitz, D.%
, Staguhn, J.%
, Armus, L.%
, Battersby, C.%
\BDBL {}others%
\end{APACrefauthors}%
\unskip\
\newblock
\APACrefYearMonthDay{2019}{}{}.
\newblock
{\BBOQ}\APACrefatitle {{Origins Space Telescope Mission Concept Study Report}}
  {{Origins Space Telescope Mission Concept Study Report}}.{\BBCQ}
\newblock
\APACjournalVolNumPages{arXiv preprint arXiv:1912.06213}{}{}{}.
\PrintBackRefs{\CurrentBib}

\bibitem [\protect \citeauthoryear {%
Des~Marais%
\ \protect \BOthers {.}}{%
Des~Marais%
\ \protect \BOthers {.}}{%
{\protect \APACyear {2008}}%
}]{%
des2008nasa}
\APACinsertmetastar {%
des2008nasa}%
\begin{APACrefauthors}%
Des~Marais, D\BPBI J.%
, Nuth~III, J\BPBI A.%
, Allamandola, L\BPBI J.%
, Boss, A\BPBI P.%
, Farmer, J\BPBI D.%
, Hoehler, T\BPBI M.%
\BDBL {}others%
\end{APACrefauthors}%
\unskip\
\newblock
\APACrefYearMonthDay{2008}{}{}.
\newblock
{\BBOQ}\APACrefatitle {The NASA astrobiology roadmap} {The nasa astrobiology
  roadmap}.{\BBCQ}
\newblock
\APACjournalVolNumPages{Astrobiology}{8}{4}{715--730}.
\PrintBackRefs{\CurrentBib}

\bibitem [\protect \citeauthoryear {%
Domagal-Goldman%
, Segura%
, Claire%
, Robinson%
\BCBL {}\ \BBA {} Meadows%
}{%
Domagal-Goldman%
\ \protect \BOthers {.}}{%
{\protect \APACyear {2014}}%
}]{%
domagal2014abiotic}
\APACinsertmetastar {%
domagal2014abiotic}%
\begin{APACrefauthors}%
Domagal-Goldman, S\BPBI D.%
, Segura, A.%
, Claire, M\BPBI W.%
, Robinson, T\BPBI D.%
\BCBL {}\ \BBA {} Meadows, V\BPBI S.%
\end{APACrefauthors}%
\unskip\
\newblock
\APACrefYearMonthDay{2014}{}{}.
\newblock
{\BBOQ}\APACrefatitle {{Abiotic ozone and oxygen in atmospheres similar to
  prebiotic Earth}} {{Abiotic ozone and oxygen in atmospheres similar to
  prebiotic Earth}}.{\BBCQ}
\newblock
\APACjournalVolNumPages{The Astrophysical Journal}{792}{2}{90}.
\PrintBackRefs{\CurrentBib}

\bibitem [\protect \citeauthoryear {%
Fauchez%
\ \protect \BOthers {.}}{%
Fauchez%
\ \protect \BOthers {.}}{%
{\protect \APACyear {2019}}%
}]{%
fauchez2019impact}
\APACinsertmetastar {%
fauchez2019impact}%
\begin{APACrefauthors}%
Fauchez, T\BPBI J.%
, Turbet, M.%
, Villanueva, G\BPBI L.%
, Wolf, E\BPBI T.%
, Arney, G.%
, Kopparapu, R\BPBI K.%
\BDBL {}others%
\end{APACrefauthors}%
\unskip\
\newblock
\APACrefYearMonthDay{2019}{}{}.
\newblock
{\BBOQ}\APACrefatitle {{Impact of clouds and hazes on the simulated JWST
  transmission spectra of habitable zone planets in the TRAPPIST-1 system}}
  {{Impact of clouds and hazes on the simulated JWST transmission spectra of
  habitable zone planets in the TRAPPIST-1 system}}.{\BBCQ}
\newblock
\APACjournalVolNumPages{The Astrophysical Journal}{887}{2}{194}.
\PrintBackRefs{\CurrentBib}

\bibitem [\protect \citeauthoryear {%
Fauchez%
\ \protect \BOthers {.}}{%
Fauchez%
\ \protect \BOthers {.}}{%
{\protect \APACyear {2020}}%
}]{%
fauchez2020sensitive}
\APACinsertmetastar {%
fauchez2020sensitive}%
\begin{APACrefauthors}%
Fauchez, T\BPBI J.%
, Villanueva, G\BPBI L.%
, Schwieterman, E\BPBI W.%
, Turbet, M.%
, Arney, G.%
, Pidhorodetska, D.%
\BDBL {}Domagal-Goldman, S\BPBI D.%
\end{APACrefauthors}%
\unskip\
\newblock
\APACrefYearMonthDay{2020}{}{}.
\newblock
{\BBOQ}\APACrefatitle {{Sensitive probing of exoplanetary oxygen via
  mid-infrared collisional absorption}} {{Sensitive probing of exoplanetary
  oxygen via mid-infrared collisional absorption}}.{\BBCQ}
\newblock
\APACjournalVolNumPages{Nature Astronomy}{}{}{1--5}.
\PrintBackRefs{\CurrentBib}

\bibitem [\protect \citeauthoryear {%
Felton%
}{%
Felton%
}{%
{\protect \APACyear {2021}}%
}]{%
feltonzenodowater2021}
\APACinsertmetastar {%
feltonzenodowater2021}%
\begin{APACrefauthors}%
Felton, R.%
\end{APACrefauthors}%
\unskip\
\newblock
\APACrefYearMonthDay{2021}{}{}.
\newblock
{\BBOQ}\APACrefatitle {Supplemental material: The Role of Atmospheric Exchange
  in False-Positive Biosignature Detection. Zenodo.
  https://doi.org/10.5281/zenodo.5591912} {Supplemental material: The role of
  atmospheric exchange in false-positive biosignature detection. zenodo.
  https://doi.org/10.5281/zenodo.5591912}.{\BBCQ}
\newblock

\PrintBackRefs{\CurrentBib}

\bibitem [\protect \citeauthoryear {%
Gao%
, Hu%
, Robinson%
, Li%
\BCBL {}\ \BBA {} Yung%
}{%
Gao%
\ \protect \BOthers {.}}{%
{\protect \APACyear {2015}}%
}]{%
gao2015stabilization}
\APACinsertmetastar {%
gao2015stabilization}%
\begin{APACrefauthors}%
Gao, P.%
, Hu, R.%
, Robinson, T\BPBI D.%
, Li, C.%
\BCBL {}\ \BBA {} Yung, Y\BPBI L.%
\end{APACrefauthors}%
\unskip\
\newblock
\APACrefYearMonthDay{2015}{}{}.
\newblock
{\BBOQ}\APACrefatitle {{Stabilization of CO2 atmospheres on exoplanets around M
  dwarf stars}} {{Stabilization of CO2 atmospheres on exoplanets around M dwarf
  stars}}.{\BBCQ}
\newblock
\APACjournalVolNumPages{Astrophys J}{806}{}{249--261}.
\PrintBackRefs{\CurrentBib}

\bibitem [\protect \citeauthoryear {%
Garcia-Sage%
, Glocer%
, Drake%
, Gronoff%
\BCBL {}\ \BBA {} Cohen%
}{%
Garcia-Sage%
\ \protect \BOthers {.}}{%
{\protect \APACyear {2017}}%
}]{%
garcia2017magnetic}
\APACinsertmetastar {%
garcia2017magnetic}%
\begin{APACrefauthors}%
Garcia-Sage, K.%
, Glocer, A.%
, Drake, J.%
, Gronoff, G.%
\BCBL {}\ \BBA {} Cohen, O.%
\end{APACrefauthors}%
\unskip\
\newblock
\APACrefYearMonthDay{2017}{}{}.
\newblock
{\BBOQ}\APACrefatitle {{On the magnetic protection of the atmosphere of Proxima
  Centauri b}} {{On the magnetic protection of the atmosphere of Proxima
  Centauri b}}.{\BBCQ}
\newblock
\APACjournalVolNumPages{The Astrophysical Journal Letters}{844}{1}{L13}.
\PrintBackRefs{\CurrentBib}

\bibitem [\protect \citeauthoryear {%
Gaudi%
\ \protect \BOthers {.}}{%
Gaudi%
\ \protect \BOthers {.}}{%
{\protect \APACyear {2020}}%
}]{%
gaudi2020habitable}
\APACinsertmetastar {%
gaudi2020habitable}%
\begin{APACrefauthors}%
Gaudi, B\BPBI S.%
, Seager, S.%
, Mennesson, B.%
, Kiessling, A.%
, Warfield, K.%
, Cahoy, K.%
\BDBL {}others%
\end{APACrefauthors}%
\unskip\
\newblock
\APACrefYearMonthDay{2020}{}{}.
\newblock
{\BBOQ}\APACrefatitle {{The Habitable Exoplanet Observatory (HabEx) Mission
  Concept Study Final Report}} {{The Habitable Exoplanet Observatory (HabEx)
  Mission Concept Study Final Report}}.{\BBCQ}
\newblock
\APACjournalVolNumPages{arXiv preprint arXiv:2001.06683}{}{}{}.
\PrintBackRefs{\CurrentBib}

\bibitem [\protect \citeauthoryear {%
Gillon%
\ \protect \BOthers {.}}{%
Gillon%
\ \protect \BOthers {.}}{%
{\protect \APACyear {2016}}%
}]{%
gillon2016temperate}
\APACinsertmetastar {%
gillon2016temperate}%
\begin{APACrefauthors}%
Gillon, M.%
, Jehin, E.%
, Lederer, S\BPBI M.%
, Delrez, L.%
, de Wit, J.%
, Burdanov, A.%
\BDBL {}others%
\end{APACrefauthors}%
\unskip\
\newblock
\APACrefYearMonthDay{2016}{}{}.
\newblock
{\BBOQ}\APACrefatitle {Temperate Earth-sized planets transiting a nearby
  ultracool dwarf star} {Temperate earth-sized planets transiting a nearby
  ultracool dwarf star}.{\BBCQ}
\newblock
\APACjournalVolNumPages{Nature}{533}{7602}{221--224}.
\PrintBackRefs{\CurrentBib}

\bibitem [\protect \citeauthoryear {%
Gillon%
\ \protect \BOthers {.}}{%
Gillon%
\ \protect \BOthers {.}}{%
{\protect \APACyear {2017}}%
}]{%
gillon2017seven}
\APACinsertmetastar {%
gillon2017seven}%
\begin{APACrefauthors}%
Gillon, M.%
, Triaud, A\BPBI H.%
, Demory, B\BHBI O.%
, Jehin, E.%
, Agol, E.%
, Deck, K\BPBI M.%
\BDBL {}others%
\end{APACrefauthors}%
\unskip\
\newblock
\APACrefYearMonthDay{2017}{}{}.
\newblock
{\BBOQ}\APACrefatitle {Seven temperate terrestrial planets around the nearby
  ultracool dwarf star TRAPPIST-1} {Seven temperate terrestrial planets around
  the nearby ultracool dwarf star trappist-1}.{\BBCQ}
\newblock
\APACjournalVolNumPages{Nature}{542}{7642}{456--460}.
\PrintBackRefs{\CurrentBib}

\bibitem [\protect \citeauthoryear {%
Gilmozzi%
\ \BBA {} Spyromilio%
}{%
Gilmozzi%
\ \BBA {} Spyromilio%
}{%
{\protect \APACyear {2007}}%
}]{%
gilmozzi2007european}
\APACinsertmetastar {%
gilmozzi2007european}%
\begin{APACrefauthors}%
Gilmozzi, R.%
\BCBT {}\ \BBA {} Spyromilio, J.%
\end{APACrefauthors}%
\unskip\
\newblock
\APACrefYearMonthDay{2007}{}{}.
\newblock
{\BBOQ}\APACrefatitle {The European extremely large telescope (E-ELT)} {The
  european extremely large telescope (e-elt)}.{\BBCQ}
\newblock
\APACjournalVolNumPages{The Messenger}{127}{11}{3}.
\PrintBackRefs{\CurrentBib}

\bibitem [\protect \citeauthoryear {%
{Gordon}%
\ \protect \BOthers {.}}{%
{Gordon}%
\ \protect \BOthers {.}}{%
{\protect \APACyear {2017}}%
}]{%
GORDON2015}
\APACinsertmetastar {%
GORDON2015}%
\begin{APACrefauthors}%
{Gordon}, I\BPBI E.%
, {Rothman}, L\BPBI S.%
, {Hill}, C.%
, {Kochanov}, R\BPBI V.%
, {Tan}, Y.%
, {Bernath}, P\BPBI F.%
\BDBL {}{Zak}, E\BPBI J.%
\end{APACrefauthors}%
\unskip\
\newblock
\APACrefYearMonthDay{2017}{{\APACmonth{12}}}{}.
\newblock
{\BBOQ}\APACrefatitle {The HITRAN2016 molecular spectroscopic database} {The
  hitran2016 molecular spectroscopic database}.{\BBCQ}
\newblock
\APACjournalVolNumPages{JQSRT}{203}{}{3-69}.
\newblock
\begin{APACrefDOI} \doi{10.1016/j.jqsrt.2017.06.038} \end{APACrefDOI}
\PrintBackRefs{\CurrentBib}

\bibitem [\protect \citeauthoryear {%
Harman%
\ \BBA {} Domagal-Goldman%
}{%
Harman%
\ \BBA {} Domagal-Goldman%
}{%
{\protect \APACyear {2018}}%
}]{%
harman2018biosignature}
\APACinsertmetastar {%
harman2018biosignature}%
\begin{APACrefauthors}%
Harman, C\BPBI E.%
\BCBT {}\ \BBA {} Domagal-Goldman, S.%
\end{APACrefauthors}%
\unskip\
\newblock
\APACrefYearMonthDay{2018}{}{}.
\newblock
{\BBOQ}\APACrefatitle {Biosignature False Positives} {Biosignature false
  positives}.{\BBCQ}
\newblock
\APACjournalVolNumPages{Handbook of Exoplanets; Springer International
  Publishing AG, part of Springer Nature: New York, NY, USA}{}{}{71}.
\PrintBackRefs{\CurrentBib}

\bibitem [\protect \citeauthoryear {%
Harman%
\ \protect \BOthers {.}}{%
Harman%
\ \protect \BOthers {.}}{%
{\protect \APACyear {2018}}%
}]{%
harman2018abiotic}
\APACinsertmetastar {%
harman2018abiotic}%
\begin{APACrefauthors}%
Harman, C\BPBI E.%
, Felton, R.%
, Hu, R.%
, Domagal-Goldman, S\BPBI D.%
, Segura, A.%
, Tian, F.%
\BCBL {}\ \BBA {} Kasting, J.%
\end{APACrefauthors}%
\unskip\
\newblock
\APACrefYearMonthDay{2018}{}{}.
\newblock
{\BBOQ}\APACrefatitle {{Abiotic O2 levels on planets around F, G, K, and M
  stars: effects of lightning-produced catalysts in eliminating oxygen false
  positives}} {{Abiotic O2 levels on planets around F, G, K, and M stars:
  effects of lightning-produced catalysts in eliminating oxygen false
  positives}}.{\BBCQ}
\newblock
\APACjournalVolNumPages{The Astrophysical Journal}{866}{1}{56}.
\PrintBackRefs{\CurrentBib}

\bibitem [\protect \citeauthoryear {%
Hartle%
\ \protect \BOthers {.}}{%
Hartle%
\ \protect \BOthers {.}}{%
{\protect \APACyear {2006}}%
}]{%
hartle2006initial}
\APACinsertmetastar {%
hartle2006initial}%
\begin{APACrefauthors}%
Hartle, R.%
, Sittler, E.%
, Neubauer, F.%
, Johnson, R.%
, Smith, H.%
, Crary, F.%
\BDBL {}others%
\end{APACrefauthors}%
\unskip\
\newblock
\APACrefYearMonthDay{2006}{}{}.
\newblock
{\BBOQ}\APACrefatitle {{Initial interpretation of Titan plasma interaction as
  observed by the Cassini plasma spectrometer: Comparisons with Voyager 1}}
  {{Initial interpretation of Titan plasma interaction as observed by the
  Cassini plasma spectrometer: Comparisons with Voyager 1}}.{\BBCQ}
\newblock
\APACjournalVolNumPages{Planetary and Space Science}{54}{12}{1211--1224}.
\PrintBackRefs{\CurrentBib}

\bibitem [\protect \citeauthoryear {%
Hartogh%
\ \protect \BOthers {.}}{%
Hartogh%
\ \protect \BOthers {.}}{%
{\protect \APACyear {2011}}%
}]{%
hartogh2011direct}
\APACinsertmetastar {%
hartogh2011direct}%
\begin{APACrefauthors}%
Hartogh, P.%
, Lellouch, E.%
, Moreno, R.%
, Bockel{\'e}e-Morvan, D.%
, Biver, N.%
, Cassidy, T.%
\BDBL {}others%
\end{APACrefauthors}%
\unskip\
\newblock
\APACrefYearMonthDay{2011}{}{}.
\newblock
{\BBOQ}\APACrefatitle {{Direct detection of the Enceladus water torus with
  Herschel}} {{Direct detection of the Enceladus water torus with
  Herschel}}.{\BBCQ}
\newblock
\APACjournalVolNumPages{Astronomy \& Astrophysics}{532}{}{L2}.
\PrintBackRefs{\CurrentBib}

\bibitem [\protect \citeauthoryear {%
H{\'e}brard%
, Dobrijevic%
, Loison%
, Bergeat%
\BCBL {}\ \BBA {} Hickson%
}{%
H{\'e}brard%
\ \protect \BOthers {.}}{%
{\protect \APACyear {2012}}%
}]{%
hebrard2012neutral}
\APACinsertmetastar {%
hebrard2012neutral}%
\begin{APACrefauthors}%
H{\'e}brard, E.%
, Dobrijevic, M.%
, Loison, J\BHBI C.%
, Bergeat, A.%
\BCBL {}\ \BBA {} Hickson, K.%
\end{APACrefauthors}%
\unskip\
\newblock
\APACrefYearMonthDay{2012}{}{}.
\newblock
{\BBOQ}\APACrefatitle {Neutral production of hydrogen isocyanide (HNC) and
  hydrogen cyanide (HCN) in Titan’s upper atmosphere} {Neutral production of
  hydrogen isocyanide (hnc) and hydrogen cyanide (hcn) in titan’s upper
  atmosphere}.{\BBCQ}
\newblock
\APACjournalVolNumPages{Astronomy \& Astrophysics}{541}{}{A21}.
\PrintBackRefs{\CurrentBib}

\bibitem [\protect \citeauthoryear {%
Hitchcock%
\ \BBA {} Lovelock%
}{%
Hitchcock%
\ \BBA {} Lovelock%
}{%
{\protect \APACyear {1967}}%
}]{%
hitchcock1967life}
\APACinsertmetastar {%
hitchcock1967life}%
\begin{APACrefauthors}%
Hitchcock, D\BPBI R.%
\BCBT {}\ \BBA {} Lovelock, J\BPBI E.%
\end{APACrefauthors}%
\unskip\
\newblock
\APACrefYearMonthDay{1967}{}{}.
\newblock
{\BBOQ}\APACrefatitle {{Life detection by atmospheric analysis}} {{Life
  detection by atmospheric analysis}}.{\BBCQ}
\newblock
\APACjournalVolNumPages{Icarus}{7}{1-3}{149--159}.
\PrintBackRefs{\CurrentBib}

\bibitem [\protect \citeauthoryear {%
H{\"o}rst%
, Vuitton%
\BCBL {}\ \BBA {} Yelle%
}{%
H{\"o}rst%
\ \protect \BOthers {.}}{%
{\protect \APACyear {2008}}%
}]{%
horst2008origin}
\APACinsertmetastar {%
horst2008origin}%
\begin{APACrefauthors}%
H{\"o}rst, S\BPBI M.%
, Vuitton, V.%
\BCBL {}\ \BBA {} Yelle, R\BPBI V.%
\end{APACrefauthors}%
\unskip\
\newblock
\APACrefYearMonthDay{2008}{}{}.
\newblock
{\BBOQ}\APACrefatitle {Origin of oxygen species in Titan's atmosphere} {Origin
  of oxygen species in titan's atmosphere}.{\BBCQ}
\newblock
\APACjournalVolNumPages{Journal of Geophysical Research: Planets}{113}{E10}{}.
\PrintBackRefs{\CurrentBib}

\bibitem [\protect \citeauthoryear {%
Johns%
}{%
Johns%
}{%
{\protect \APACyear {2006}}%
}]{%
johns2006giant}
\APACinsertmetastar {%
johns2006giant}%
\begin{APACrefauthors}%
Johns, M.%
\end{APACrefauthors}%
\unskip\
\newblock
\APACrefYearMonthDay{2006}{}{}.
\newblock
{\BBOQ}\APACrefatitle {The giant Magellan telescope (GMT)} {The giant magellan
  telescope (gmt)}.{\BBCQ}
\newblock
\BIn{} \APACrefbtitle {Ground-based and Airborne Telescopes} {Ground-based and
  airborne telescopes}\ (\BVOL\ 6267, \BPG~626729).
\PrintBackRefs{\CurrentBib}

\bibitem [\protect \citeauthoryear {%
Kasting%
\ \BBA {} Ackerman%
}{%
Kasting%
\ \BBA {} Ackerman%
}{%
{\protect \APACyear {1986}}%
}]{%
kasting1986climatic}
\APACinsertmetastar {%
kasting1986climatic}%
\begin{APACrefauthors}%
Kasting, J\BPBI F.%
\BCBT {}\ \BBA {} Ackerman, T\BPBI P.%
\end{APACrefauthors}%
\unskip\
\newblock
\APACrefYearMonthDay{1986}{}{}.
\newblock
{\BBOQ}\APACrefatitle {Climatic consequences of very high carbon dioxide levels
  in the Earth's early atmosphere} {Climatic consequences of very high carbon
  dioxide levels in the earth's early atmosphere}.{\BBCQ}
\newblock
\APACjournalVolNumPages{Science}{234}{4782}{1383--1385}.
\PrintBackRefs{\CurrentBib}

\bibitem [\protect \citeauthoryear {%
Kasting%
, Whitmire%
\BCBL {}\ \BBA {} Reynolds%
}{%
Kasting%
\ \protect \BOthers {.}}{%
{\protect \APACyear {1993}}%
}]{%
kasting1993habitable}
\APACinsertmetastar {%
kasting1993habitable}%
\begin{APACrefauthors}%
Kasting, J\BPBI F.%
, Whitmire, D\BPBI P.%
\BCBL {}\ \BBA {} Reynolds, R\BPBI T.%
\end{APACrefauthors}%
\unskip\
\newblock
\APACrefYearMonthDay{1993}{}{}.
\newblock
{\BBOQ}\APACrefatitle {Habitable zones around main sequence stars} {Habitable
  zones around main sequence stars}.{\BBCQ}
\newblock
\APACjournalVolNumPages{Icarus}{101}{1}{108--128}.
\PrintBackRefs{\CurrentBib}

\bibitem [\protect \citeauthoryear {%
Kharecha%
, Kasting%
\BCBL {}\ \BBA {} Siefert%
}{%
Kharecha%
\ \protect \BOthers {.}}{%
{\protect \APACyear {2005}}%
}]{%
Kharecha2005}
\APACinsertmetastar {%
Kharecha2005}%
\begin{APACrefauthors}%
Kharecha, P.%
, Kasting, J.%
\BCBL {}\ \BBA {} Siefert, J.%
\end{APACrefauthors}%
\unskip\
\newblock
\APACrefYearMonthDay{2005}{}{}.
\newblock
{\BBOQ}\APACrefatitle {A coupled atmosphere–ecosystem model of the early
  Archean Earth} {A coupled atmosphere–ecosystem model of the early archean
  earth}.{\BBCQ}
\newblock
\APACjournalVolNumPages{Geobiology}{3}{2}{53-76}.
\newblock
\begin{APACrefURL}
  \url{https://onlinelibrary.wiley.com/doi/abs/10.1111/j.1472-4669.2005.00049.x}
  \end{APACrefURL}
\newblock
\begin{APACrefDOI} \doi{10.1111/j.1472-4669.2005.00049.x} \end{APACrefDOI}
\PrintBackRefs{\CurrentBib}

\bibitem [\protect \citeauthoryear {%
{Kopparapu}%
\ \protect \BOthers {.}}{%
{Kopparapu}%
\ \protect \BOthers {.}}{%
{\protect \APACyear {2013}}%
}]{%
Kopparapu2013}
\APACinsertmetastar {%
Kopparapu2013}%
\begin{APACrefauthors}%
{Kopparapu}, R\BPBI K.%
, {Ramirez}, R.%
, {Kasting}, J\BPBI F.%
, {Eymet}, V.%
, {Robinson}, T\BPBI D.%
, {Mahadevan}, S.%
\BDBL {}{Deshpande}, R.%
\end{APACrefauthors}%
\unskip\
\newblock
\APACrefYearMonthDay{2013}{{\APACmonth{03}}}{}.
\newblock
{\BBOQ}\APACrefatitle {Habitable Zones around Main-sequence Stars: New
  Estimates} {Habitable zones around main-sequence stars: New
  estimates}.{\BBCQ}
\newblock
\APACjournalVolNumPages{The Astrophysical Journal}{765}{}{131}.
\newblock
\begin{APACrefDOI} \doi{10.1088/0004-637X/765/2/131} \end{APACrefDOI}
\PrintBackRefs{\CurrentBib}

\bibitem [\protect \citeauthoryear {%
Krasnopolsky%
}{%
Krasnopolsky%
}{%
{\protect \APACyear {2009}}%
}]{%
krasnopolsky2009photochemical}
\APACinsertmetastar {%
krasnopolsky2009photochemical}%
\begin{APACrefauthors}%
Krasnopolsky, V\BPBI A.%
\end{APACrefauthors}%
\unskip\
\newblock
\APACrefYearMonthDay{2009}{}{}.
\newblock
{\BBOQ}\APACrefatitle {{A photochemical model of Titan's atmosphere and
  ionosphere}} {{A photochemical model of Titan's atmosphere and
  ionosphere}}.{\BBCQ}
\newblock
\APACjournalVolNumPages{Icarus}{201}{1}{226--256}.
\PrintBackRefs{\CurrentBib}

\bibitem [\protect \citeauthoryear {%
Krissansen-Totton%
, Bergsman%
\BCBL {}\ \BBA {} Catling%
}{%
Krissansen-Totton%
\ \protect \BOthers {.}}{%
{\protect \APACyear {2016}}%
}]{%
krissansen2016detecting}
\APACinsertmetastar {%
krissansen2016detecting}%
\begin{APACrefauthors}%
Krissansen-Totton, J.%
, Bergsman, D\BPBI S.%
\BCBL {}\ \BBA {} Catling, D\BPBI C.%
\end{APACrefauthors}%
\unskip\
\newblock
\APACrefYearMonthDay{2016}{}{}.
\newblock
{\BBOQ}\APACrefatitle {On detecting biospheres from chemical thermodynamic
  disequilibrium in planetary atmospheres} {On detecting biospheres from
  chemical thermodynamic disequilibrium in planetary atmospheres}.{\BBCQ}
\newblock
\APACjournalVolNumPages{Astrobiology}{16}{1}{39--67}.
\PrintBackRefs{\CurrentBib}

\bibitem [\protect \citeauthoryear {%
Lederberg%
}{%
Lederberg%
}{%
{\protect \APACyear {1965}}%
}]{%
lederberg1965signs}
\APACinsertmetastar {%
lederberg1965signs}%
\begin{APACrefauthors}%
Lederberg, J.%
\end{APACrefauthors}%
\unskip\
\newblock
\APACrefYearMonthDay{1965}{}{}.
\newblock
{\BBOQ}\APACrefatitle {Signs of life: criterion-system of exobiology} {Signs of
  life: criterion-system of exobiology}.{\BBCQ}
\newblock
\APACjournalVolNumPages{Nature}{207}{4992}{9--13}.
\PrintBackRefs{\CurrentBib}

\bibitem [\protect \citeauthoryear {%
Levy%
}{%
Levy%
}{%
{\protect \APACyear {1971}}%
}]{%
levy1971normal}
\APACinsertmetastar {%
levy1971normal}%
\begin{APACrefauthors}%
Levy, H.%
\end{APACrefauthors}%
\unskip\
\newblock
\APACrefYearMonthDay{1971}{}{}.
\newblock
{\BBOQ}\APACrefatitle {Normal atmosphere: Large radical and formaldehyde
  concentrations predicted} {Normal atmosphere: Large radical and formaldehyde
  concentrations predicted}.{\BBCQ}
\newblock
\APACjournalVolNumPages{Science}{173}{3992}{141--143}.
\PrintBackRefs{\CurrentBib}

\bibitem [\protect \citeauthoryear {%
Lincowski%
\ \protect \BOthers {.}}{%
Lincowski%
\ \protect \BOthers {.}}{%
{\protect \APACyear {2018}}%
}]{%
lincowski2018evolved}
\APACinsertmetastar {%
lincowski2018evolved}%
\begin{APACrefauthors}%
Lincowski, A\BPBI P.%
, Meadows, V\BPBI S.%
, Crisp, D.%
, Robinson, T\BPBI D.%
, Luger, R.%
, Lustig-Yaeger, J.%
\BCBL {}\ \BBA {} Arney, G\BPBI N.%
\end{APACrefauthors}%
\unskip\
\newblock
\APACrefYearMonthDay{2018}{}{}.
\newblock
{\BBOQ}\APACrefatitle {Evolved climates and observational discriminants for the
  TRAPPIST-1 planetary system} {Evolved climates and observational
  discriminants for the trappist-1 planetary system}.{\BBCQ}
\newblock
\APACjournalVolNumPages{The Astrophysical Journal}{867}{1}{76}.
\PrintBackRefs{\CurrentBib}

\bibitem [\protect \citeauthoryear {%
Lovelock%
}{%
Lovelock%
}{%
{\protect \APACyear {1965}}%
}]{%
lovelock1965physical}
\APACinsertmetastar {%
lovelock1965physical}%
\begin{APACrefauthors}%
Lovelock, J\BPBI E.%
\end{APACrefauthors}%
\unskip\
\newblock
\APACrefYearMonthDay{1965}{}{}.
\newblock
{\BBOQ}\APACrefatitle {A physical basis for life detection experiments} {A
  physical basis for life detection experiments}.{\BBCQ}
\newblock
\APACjournalVolNumPages{Nature}{207}{4997}{568--570}.
\PrintBackRefs{\CurrentBib}

\bibitem [\protect \citeauthoryear {%
Lovelock%
}{%
Lovelock%
}{%
{\protect \APACyear {1975}}%
}]{%
lovelock1975thermodynamics}
\APACinsertmetastar {%
lovelock1975thermodynamics}%
\begin{APACrefauthors}%
Lovelock, J\BPBI E.%
\end{APACrefauthors}%
\unskip\
\newblock
\APACrefYearMonthDay{1975}{}{}.
\newblock
{\BBOQ}\APACrefatitle {{Thermodynamics and the recognition of alien
  biospheres}} {{Thermodynamics and the recognition of alien
  biospheres}}.{\BBCQ}
\newblock
\APACjournalVolNumPages{Proceedings of the Royal Society of London. Series B.
  Biological Sciences}{189}{1095}{167--181}.
\PrintBackRefs{\CurrentBib}

\bibitem [\protect \citeauthoryear {%
Luger%
\ \BBA {} Barnes%
}{%
Luger%
\ \BBA {} Barnes%
}{%
{\protect \APACyear {2015}}%
}]{%
luger2015extreme}
\APACinsertmetastar {%
luger2015extreme}%
\begin{APACrefauthors}%
Luger, R.%
\BCBT {}\ \BBA {} Barnes, R.%
\end{APACrefauthors}%
\unskip\
\newblock
\APACrefYearMonthDay{2015}{}{}.
\newblock
{\BBOQ}\APACrefatitle {{Extreme water loss and abiotic O2 buildup on planets
  throughout the habitable zones of M dwarfs}} {{Extreme water loss and abiotic
  O2 buildup on planets throughout the habitable zones of M dwarfs}}.{\BBCQ}
\newblock
\APACjournalVolNumPages{Astrobiology}{15}{2}{119--143}.
\PrintBackRefs{\CurrentBib}

\bibitem [\protect \citeauthoryear {%
LUVOIR%
}{%
LUVOIR%
}{%
{\protect \APACyear {2019}}%
}]{%
luvoir2019luvoir}
\APACinsertmetastar {%
luvoir2019luvoir}%
\begin{APACrefauthors}%
LUVOIR.%
\end{APACrefauthors}%
\unskip\
\newblock
\APACrefYearMonthDay{2019}{}{}.
\newblock
{\BBOQ}\APACrefatitle {{The LUVOIR Mission Concept Study Final Report}} {{The
  LUVOIR Mission Concept Study Final Report}}.{\BBCQ}
\newblock
\APACjournalVolNumPages{arXiv preprint arXiv:1912.06219}{}{}{}.
\PrintBackRefs{\CurrentBib}

\bibitem [\protect \citeauthoryear {%
{Manabe}%
\ \BBA {} {Wetherald}%
}{%
{Manabe}%
\ \BBA {} {Wetherald}%
}{%
{\protect \APACyear {1967}}%
}]{%
Manabe1967}
\APACinsertmetastar {%
Manabe1967}%
\begin{APACrefauthors}%
{Manabe}, S.%
\BCBT {}\ \BBA {} {Wetherald}, R\BPBI T.%
\end{APACrefauthors}%
\unskip\
\newblock
\APACrefYearMonthDay{1967}{{\APACmonth{05}}}{}.
\newblock
{\BBOQ}\APACrefatitle {{Thermal Equilibrium of the Atmosphere with a Given
  Distribution of Relative Humidity.}} {{Thermal Equilibrium of the Atmosphere
  with a Given Distribution of Relative Humidity.}}{\BBCQ}
\newblock
\APACjournalVolNumPages{Journal of Atmospheric Sciences}{24}{3}{241-259}.
\newblock
\begin{APACrefDOI} \doi{10.1175/1520-0469(1967)024<0241:TEOTAW>2.0.CO;2}
  \end{APACrefDOI}
\PrintBackRefs{\CurrentBib}

\bibitem [\protect \citeauthoryear {%
Marais%
\ \BBA {} Walter%
}{%
Marais%
\ \BBA {} Walter%
}{%
{\protect \APACyear {1999}}%
}]{%
marais1999astrobiology}
\APACinsertmetastar {%
marais1999astrobiology}%
\begin{APACrefauthors}%
Marais, D\BPBI D.%
\BCBT {}\ \BBA {} Walter, M.%
\end{APACrefauthors}%
\unskip\
\newblock
\APACrefYearMonthDay{1999}{}{}.
\newblock
{\BBOQ}\APACrefatitle {Astrobiology: exploring the origins, evolution, and
  distribution of life in the universe} {Astrobiology: exploring the origins,
  evolution, and distribution of life in the universe}.{\BBCQ}
\newblock
\APACjournalVolNumPages{Annual review of ecology and
  systematics}{30}{1}{397--420}.
\PrintBackRefs{\CurrentBib}

\bibitem [\protect \citeauthoryear {%
Meadows%
}{%
Meadows%
}{%
{\protect \APACyear {2017}}%
}]{%
meadows2017reflections}
\APACinsertmetastar {%
meadows2017reflections}%
\begin{APACrefauthors}%
Meadows, V\BPBI S.%
\end{APACrefauthors}%
\unskip\
\newblock
\APACrefYearMonthDay{2017}{}{}.
\newblock
{\BBOQ}\APACrefatitle {{Reflections on O2 as a biosignature in exoplanetary
  atmospheres}} {{Reflections on O2 as a biosignature in exoplanetary
  atmospheres}}.{\BBCQ}
\newblock
\APACjournalVolNumPages{Astrobiology}{17}{10}{1022--1052}.
\PrintBackRefs{\CurrentBib}

\bibitem [\protect \citeauthoryear {%
Meadows%
\ \protect \BOthers {.}}{%
Meadows%
\ \protect \BOthers {.}}{%
{\protect \APACyear {2018}}%
}]{%
meadows2018habitability}
\APACinsertmetastar {%
meadows2018habitability}%
\begin{APACrefauthors}%
Meadows, V\BPBI S.%
, Arney, G\BPBI N.%
, Schwieterman, E\BPBI W.%
, Lustig-Yaeger, J.%
, Lincowski, A\BPBI P.%
, Robinson, T.%
\BDBL {}others%
\end{APACrefauthors}%
\unskip\
\newblock
\APACrefYearMonthDay{2018}{}{}.
\newblock
{\BBOQ}\APACrefatitle {The habitability of Proxima Centauri b: environmental
  states and observational discriminants} {The habitability of proxima centauri
  b: environmental states and observational discriminants}.{\BBCQ}
\newblock
\APACjournalVolNumPages{Astrobiology}{18}{2}{133--189}.
\PrintBackRefs{\CurrentBib}

\bibitem [\protect \citeauthoryear {%
Moreno%
\ \protect \BOthers {.}}{%
Moreno%
\ \protect \BOthers {.}}{%
{\protect \APACyear {2012}}%
}]{%
moreno2012abundance}
\APACinsertmetastar {%
moreno2012abundance}%
\begin{APACrefauthors}%
Moreno, R.%
, Lellouch, E.%
, Lara, L\BPBI M.%
, Feuchtgruber, H.%
, Rengel, M.%
, Hartogh, P.%
\BCBL {}\ \BBA {} Courtin, R.%
\end{APACrefauthors}%
\unskip\
\newblock
\APACrefYearMonthDay{2012}{}{}.
\newblock
{\BBOQ}\APACrefatitle {{The abundance, vertical distribution and origin of H2O
  in Titan’s atmosphere: Herschel observations and photochemical modelling}}
  {{The abundance, vertical distribution and origin of H2O in Titan’s
  atmosphere: Herschel observations and photochemical modelling}}.{\BBCQ}
\newblock
\APACjournalVolNumPages{Icarus}{221}{2}{753--767}.
\PrintBackRefs{\CurrentBib}

\bibitem [\protect \citeauthoryear {%
National Academies~of Sciences%
\ \BBA {} Medicine%
}{%
National Academies~of Sciences%
\ \BBA {} Medicine%
}{%
{\protect \APACyear {2021}}%
}]{%
national2021pathways}
\APACinsertmetastar {%
national2021pathways}%
\begin{APACrefauthors}%
National Academies~of Sciences, E.%
\BCBT {}\ \BBA {} Medicine.%
\end{APACrefauthors}%
\unskip\
\newblock
\APACrefYearMonthDay{2021}{}{}.
\newblock
{\BBOQ}\APACrefatitle {Pathways to Discovery in Astronomy and Astrophysics for
  the 2020s} {Pathways to discovery in astronomy and astrophysics for the
  2020s}.{\BBCQ}
\newblock
\APACjournalVolNumPages{Washington DC: The National Academies Press.
  https://doi.org/10.17226/26141}{}{}{}.
\PrintBackRefs{\CurrentBib}

\bibitem [\protect \citeauthoryear {%
Niemann%
\ \protect \BOthers {.}}{%
Niemann%
\ \protect \BOthers {.}}{%
{\protect \APACyear {2010}}%
}]{%
niemann2010composition}
\APACinsertmetastar {%
niemann2010composition}%
\begin{APACrefauthors}%
Niemann, H.%
, Atreya, S.%
, Demick, J.%
, Gautier, D.%
, Haberman, J.%
, Harpold, D.%
\BDBL {}Raulin, F.%
\end{APACrefauthors}%
\unskip\
\newblock
\APACrefYearMonthDay{2010}{}{}.
\newblock
{\BBOQ}\APACrefatitle {Composition of Titan's lower atmosphere and simple
  surface volatiles as measured by the Cassini-Huygens probe gas chromatograph
  mass spectrometer experiment} {Composition of titan's lower atmosphere and
  simple surface volatiles as measured by the cassini-huygens probe gas
  chromatograph mass spectrometer experiment}.{\BBCQ}
\newblock
\APACjournalVolNumPages{Journal of Geophysical Research: Planets}{115}{E12}{}.
\PrintBackRefs{\CurrentBib}

\bibitem [\protect \citeauthoryear {%
Pavlov%
, Brown%
\BCBL {}\ \BBA {} Kasting%
}{%
Pavlov%
\ \protect \BOthers {.}}{%
{\protect \APACyear {2001}}%
}]{%
Pavlov2001a}
\APACinsertmetastar {%
Pavlov2001a}%
\begin{APACrefauthors}%
Pavlov, A\BPBI A.%
, Brown, L\BPBI L.%
\BCBL {}\ \BBA {} Kasting, J\BPBI F.%
\end{APACrefauthors}%
\unskip\
\newblock
\APACrefYearMonthDay{2001}{}{}.
\newblock
{\BBOQ}\APACrefatitle {{UV} shielding of {NH$_\textnormal{3}$} and
  {O$_\textnormal{2}$} by organic hazes in the {Archean} atmosphere} {{UV}
  shielding of {NH$_\textnormal{3}$} and {O$_\textnormal{2}$} by organic hazes
  in the {Archean} atmosphere}.{\BBCQ}
\newblock
\APACjournalVolNumPages{Journal of Geophysical Research:
  Planets}{106}{E10}{23267-23287}.
\newblock
\begin{APACrefURL}
  \url{https://agupubs.onlinelibrary.wiley.com/doi/abs/10.1029/2000JE001448}
  \end{APACrefURL}
\newblock
\begin{APACrefDOI} \doi{10.1029/2000JE001448} \end{APACrefDOI}
\PrintBackRefs{\CurrentBib}

\bibitem [\protect \citeauthoryear {%
Peacock%
, Barman%
, Shkolnik%
, Hauschildt%
\BCBL {}\ \BBA {} Baron%
}{%
Peacock%
\ \protect \BOthers {.}}{%
{\protect \APACyear {2019}}%
}]{%
peacock2019predicting}
\APACinsertmetastar {%
peacock2019predicting}%
\begin{APACrefauthors}%
Peacock, S.%
, Barman, T.%
, Shkolnik, E\BPBI L.%
, Hauschildt, P\BPBI H.%
\BCBL {}\ \BBA {} Baron, E.%
\end{APACrefauthors}%
\unskip\
\newblock
\APACrefYearMonthDay{2019}{}{}.
\newblock
{\BBOQ}\APACrefatitle {Predicting the extreme ultraviolet radiation environment
  of exoplanets around Low-mass stars: the TRAPPIST-1 system} {Predicting the
  extreme ultraviolet radiation environment of exoplanets around low-mass
  stars: the trappist-1 system}.{\BBCQ}
\newblock
\APACjournalVolNumPages{The Astrophysical Journal}{871}{2}{235}.
\PrintBackRefs{\CurrentBib}

\bibitem [\protect \citeauthoryear {%
{Pidhorodetska}%
, {Fauchez}%
, {Villanueva}%
, {Domagal-Goldman}%
\BCBL {}\ \BBA {} {Kopparapu}%
}{%
{Pidhorodetska}%
\ \protect \BOthers {.}}{%
{\protect \APACyear {2020}}%
}]{%
Pidhorodetska2020}
\APACinsertmetastar {%
Pidhorodetska2020}%
\begin{APACrefauthors}%
{Pidhorodetska}, D.%
, {Fauchez}, T\BPBI J.%
, {Villanueva}, G\BPBI L.%
, {Domagal-Goldman}, S\BPBI D.%
\BCBL {}\ \BBA {} {Kopparapu}, R\BPBI K.%
\end{APACrefauthors}%
\unskip\
\newblock
\APACrefYearMonthDay{2020}{{\APACmonth{08}}}{}.
\newblock
{\BBOQ}\APACrefatitle {{Detectability of Molecular Signatures on TRAPPIST-1e
  through Transmission Spectroscopy Simulated for Future Space-based
  Observatories}} {{Detectability of Molecular Signatures on TRAPPIST-1e
  through Transmission Spectroscopy Simulated for Future Space-based
  Observatories}}.{\BBCQ}
\newblock
\APACjournalVolNumPages{The Astrophysical Journal Letters}{898}{2}{L33}.
\newblock
\begin{APACrefDOI} \doi{10.3847/2041-8213/aba4a1} \end{APACrefDOI}
\PrintBackRefs{\CurrentBib}

\bibitem [\protect \citeauthoryear {%
Porco%
\ \protect \BOthers {.}}{%
Porco%
\ \protect \BOthers {.}}{%
{\protect \APACyear {2006}}%
}]{%
porco2006cassini}
\APACinsertmetastar {%
porco2006cassini}%
\begin{APACrefauthors}%
Porco, C\BPBI C.%
, Helfenstein, P.%
, Thomas, P.%
, Ingersoll, A.%
, Wisdom, J.%
, West, R.%
\BDBL {}others%
\end{APACrefauthors}%
\unskip\
\newblock
\APACrefYearMonthDay{2006}{}{}.
\newblock
{\BBOQ}\APACrefatitle {{Cassini observes the active south pole of Enceladus}}
  {{Cassini observes the active south pole of Enceladus}}.{\BBCQ}
\newblock
\APACjournalVolNumPages{science}{311}{5766}{1393--1401}.
\PrintBackRefs{\CurrentBib}

\bibitem [\protect \citeauthoryear {%
Schwieterman%
\ \protect \BOthers {.}}{%
Schwieterman%
\ \protect \BOthers {.}}{%
{\protect \APACyear {2018}}%
}]{%
schwieterman2018exoplanet}
\APACinsertmetastar {%
schwieterman2018exoplanet}%
\begin{APACrefauthors}%
Schwieterman, E\BPBI W.%
, Kiang, N\BPBI Y.%
, Parenteau, M\BPBI N.%
, Harman, C\BPBI E.%
, DasSarma, S.%
, Fisher, T\BPBI M.%
\BDBL {}others%
\end{APACrefauthors}%
\unskip\
\newblock
\APACrefYearMonthDay{2018}{}{}.
\newblock
{\BBOQ}\APACrefatitle {Exoplanet biosignatures: a review of remotely detectable
  signs of life} {Exoplanet biosignatures: a review of remotely detectable
  signs of life}.{\BBCQ}
\newblock
\APACjournalVolNumPages{Astrobiology}{18}{6}{663--708}.
\PrintBackRefs{\CurrentBib}

\bibitem [\protect \citeauthoryear {%
Sittler%
\ \protect \BOthers {.}}{%
Sittler%
\ \protect \BOthers {.}}{%
{\protect \APACyear {2009}}%
}]{%
sittler2009heavy}
\APACinsertmetastar {%
sittler2009heavy}%
\begin{APACrefauthors}%
Sittler, E.%
, Ali, A.%
, Cooper, J.%
, Hartle, R.%
, Johnson, R.%
, Coates, A.%
\BCBL {}\ \BBA {} Young, D.%
\end{APACrefauthors}%
\unskip\
\newblock
\APACrefYearMonthDay{2009}{}{}.
\newblock
{\BBOQ}\APACrefatitle {Heavy ion formation in Titan's ionosphere:
  Magnetospheric introduction of free oxygen and a source of Titan's aerosols?}
  {Heavy ion formation in titan's ionosphere: Magnetospheric introduction of
  free oxygen and a source of titan's aerosols?}{\BBCQ}
\newblock
\APACjournalVolNumPages{Planetary and Space Science}{57}{13}{1547--1557}.
\PrintBackRefs{\CurrentBib}

\bibitem [\protect \citeauthoryear {%
Skidmore%
\ \protect \BOthers {.}}{%
Skidmore%
\ \protect \BOthers {.}}{%
{\protect \APACyear {2015}}%
}]{%
skidmore2015thirty}
\APACinsertmetastar {%
skidmore2015thirty}%
\begin{APACrefauthors}%
Skidmore, W.%
\BCBT {}\ \BOthersPeriod {.}
\end{APACrefauthors}%
\unskip\
\newblock
\APACrefYearMonthDay{2015}{}{}.
\newblock
{\BBOQ}\APACrefatitle {Thirty meter telescope detailed science case: 2015}
  {Thirty meter telescope detailed science case: 2015}.{\BBCQ}
\newblock
\APACjournalVolNumPages{Research in Astronomy and Astrophysics}{15}{12}{1945}.
\PrintBackRefs{\CurrentBib}

\bibitem [\protect \citeauthoryear {%
{Toon}%
, {McKay}%
, {Ackerman}%
\BCBL {}\ \BBA {} {Santhanam}%
}{%
{Toon}%
\ \protect \BOthers {.}}{%
{\protect \APACyear {1989}}%
}]{%
Toon1989}
\APACinsertmetastar {%
Toon1989}%
\begin{APACrefauthors}%
{Toon}, O\BPBI B.%
, {McKay}, C\BPBI P.%
, {Ackerman}, T\BPBI P.%
\BCBL {}\ \BBA {} {Santhanam}, K.%
\end{APACrefauthors}%
\unskip\
\newblock
\APACrefYearMonthDay{1989}{{\APACmonth{11}}}{}.
\newblock
{\BBOQ}\APACrefatitle {{Rapid calculation of radiative heating rates and
  photodissociation rates in inhomogeneous multiple scattering atmospheres}}
  {{Rapid calculation of radiative heating rates and photodissociation rates in
  inhomogeneous multiple scattering atmospheres}}.{\BBCQ}
\newblock
\APACjournalVolNumPages{Journal of Geophysical Research}{94}{}{16287-16301}.
\newblock
\begin{APACrefDOI} \doi{10.1029/JD094iD13p16287} \end{APACrefDOI}
\PrintBackRefs{\CurrentBib}

\bibitem [\protect \citeauthoryear {%
Trainer%
\ \protect \BOthers {.}}{%
Trainer%
\ \protect \BOthers {.}}{%
{\protect \APACyear {2006}}%
}]{%
trainer2006organic}
\APACinsertmetastar {%
trainer2006organic}%
\begin{APACrefauthors}%
Trainer, M\BPBI G.%
, Pavlov, A\BPBI A.%
, DeWitt, H\BPBI L.%
, Jimenez, J\BPBI L.%
, McKay, C\BPBI P.%
, Toon, O\BPBI B.%
\BCBL {}\ \BBA {} Tolbert, M\BPBI A.%
\end{APACrefauthors}%
\unskip\
\newblock
\APACrefYearMonthDay{2006}{}{}.
\newblock
{\BBOQ}\APACrefatitle {Organic haze on Titan and the early Earth} {Organic haze
  on titan and the early earth}.{\BBCQ}
\newblock
\APACjournalVolNumPages{Proceedings of the National Academy of
  Sciences}{103}{48}{18035--18042}.
\PrintBackRefs{\CurrentBib}

\bibitem [\protect \citeauthoryear {%
Ueno%
, Yamada%
, Yoshida%
, Maruyama%
\BCBL {}\ \BBA {} Isozaki%
}{%
Ueno%
\ \protect \BOthers {.}}{%
{\protect \APACyear {2006}}%
}]{%
ueno2006evidence}
\APACinsertmetastar {%
ueno2006evidence}%
\begin{APACrefauthors}%
Ueno, Y.%
, Yamada, K.%
, Yoshida, N.%
, Maruyama, S.%
\BCBL {}\ \BBA {} Isozaki, Y.%
\end{APACrefauthors}%
\unskip\
\newblock
\APACrefYearMonthDay{2006}{}{}.
\newblock
{\BBOQ}\APACrefatitle {Evidence from fluid inclusions for microbial
  methanogenesis in the early Archaean era} {Evidence from fluid inclusions for
  microbial methanogenesis in the early archaean era}.{\BBCQ}
\newblock
\APACjournalVolNumPages{Nature}{440}{7083}{516--519}.
\PrintBackRefs{\CurrentBib}

\bibitem [\protect \citeauthoryear {%
Villanueva%
, Smith%
, Protopapa%
, Faggi%
\BCBL {}\ \BBA {} Mandell%
}{%
Villanueva%
\ \protect \BOthers {.}}{%
{\protect \APACyear {2018}}%
}]{%
villanueva2018planetary}
\APACinsertmetastar {%
villanueva2018planetary}%
\begin{APACrefauthors}%
Villanueva, G\BPBI L.%
, Smith, M\BPBI D.%
, Protopapa, S.%
, Faggi, S.%
\BCBL {}\ \BBA {} Mandell, A\BPBI M.%
\end{APACrefauthors}%
\unskip\
\newblock
\APACrefYearMonthDay{2018}{}{}.
\newblock
{\BBOQ}\APACrefatitle {Planetary Spectrum Generator: An accurate online
  radiative transfer suite for atmospheres, comets, small bodies and
  exoplanets} {Planetary spectrum generator: An accurate online radiative
  transfer suite for atmospheres, comets, small bodies and exoplanets}.{\BBCQ}
\newblock
\APACjournalVolNumPages{Journal of Quantitative Spectroscopy and Radiative
  Transfer}{217}{}{86--104}.
\PrintBackRefs{\CurrentBib}

\bibitem [\protect \citeauthoryear {%
Wagner%
\ \protect \BOthers {.}}{%
Wagner%
\ \protect \BOthers {.}}{%
{\protect \APACyear {2021}}%
}]{%
wagner2021imaging}
\APACinsertmetastar {%
wagner2021imaging}%
\begin{APACrefauthors}%
Wagner, K.%
, Boehle, A.%
, Pathak, P.%
, Kasper, M.%
, Arsenault, R.%
, Jakob, G.%
\BDBL {}others%
\end{APACrefauthors}%
\unskip\
\newblock
\APACrefYearMonthDay{2021}{}{}.
\newblock
{\BBOQ}\APACrefatitle {Imaging low-mass planets within the habitable zone of
  $\alpha$ Centauri} {Imaging low-mass planets within the habitable zone of
  $\alpha$ centauri}.{\BBCQ}
\newblock
\APACjournalVolNumPages{Nature communications}{12}{1}{1--7}.
\PrintBackRefs{\CurrentBib}

\bibitem [\protect \citeauthoryear {%
Walker%
}{%
Walker%
}{%
{\protect \APACyear {1974}}%
}]{%
walker1974stability}
\APACinsertmetastar {%
walker1974stability}%
\begin{APACrefauthors}%
Walker, J\BPBI C.%
\end{APACrefauthors}%
\unskip\
\newblock
\APACrefYearMonthDay{1974}{}{}.
\newblock
{\BBOQ}\APACrefatitle {Stability of atmospheric oxygen} {Stability of
  atmospheric oxygen}.{\BBCQ}
\newblock
\APACjournalVolNumPages{American Journal of Science}{274}{3}{193--214}.
\PrintBackRefs{\CurrentBib}

\bibitem [\protect \citeauthoryear {%
Wilson%
\ \protect \BOthers {.}}{%
Wilson%
\ \protect \BOthers {.}}{%
{\protect \APACyear {2021}}%
}]{%
wilson2021mega}
\APACinsertmetastar {%
wilson2021mega}%
\begin{APACrefauthors}%
Wilson, D\BPBI J.%
, Froning, C\BPBI S.%
, Duvvuri, G\BPBI M.%
, France, K.%
, Youngblood, A.%
, Schneider, P\BPBI C.%
\BDBL {}others%
\end{APACrefauthors}%
\unskip\
\newblock
\APACrefYearMonthDay{2021}{}{}.
\newblock
{\BBOQ}\APACrefatitle {The Mega-MUSCLES Spectral Energy Distribution of
  TRAPPIST-1} {The mega-muscles spectral energy distribution of
  trappist-1}.{\BBCQ}
\newblock
\APACjournalVolNumPages{The Astrophysical Journal}{911}{1}{18}.
\PrintBackRefs{\CurrentBib}

\bibitem [\protect \citeauthoryear {%
Wordsworth%
\ \protect \BOthers {.}}{%
Wordsworth%
\ \protect \BOthers {.}}{%
{\protect \APACyear {2011}}%
}]{%
wordsworth2011gliese}
\APACinsertmetastar {%
wordsworth2011gliese}%
\begin{APACrefauthors}%
Wordsworth, R\BPBI D.%
, Forget, F.%
, Selsis, F.%
, Millour, E.%
, Charnay, B.%
\BCBL {}\ \BBA {} Madeleine, J\BHBI B.%
\end{APACrefauthors}%
\unskip\
\newblock
\APACrefYearMonthDay{2011}{}{}.
\newblock
{\BBOQ}\APACrefatitle {Gliese 581d is the first discovered terrestrial-mass
  exoplanet in the habitable zone} {Gliese 581d is the first discovered
  terrestrial-mass exoplanet in the habitable zone}.{\BBCQ}
\newblock
\APACjournalVolNumPages{The Astrophysical Journal Letters}{733}{2}{L48}.
\PrintBackRefs{\CurrentBib}

\end{thebibliography}

\end{document}